\documentclass[11pt]{article}
\usepackage{amsfonts}
\usepackage{epic}
\usepackage{eepic}
\def\cD{{\cal D}}

\def\cP{{\cal P}}                    
          \def\cT{{\cal T}}

\def\ZZ{{\mathbb Z}}

\newcommand{\si}{\sigma}

\newcommand{\pr}{\prime}
\newcommand{\nn}{\nonumber}

\begin{document}
\pagestyle{empty}

\begin{center}
{\large{\bf On the theory of plateau-plateau transitions in
Quantum Hall Effect }}\\
 \vspace{36pt}
 {\bf A.~Sedrakyan\footnote{e-mail:{\sl
sedrak@lx2.yerphi.am, sedrak@alf.nbi.dk}}}\\

\vspace{30pt}
\emph{Yerevan Physics Institute,}
\\
\emph{Br.Alikhanian str.2, Yerevan 36, Armenia}
\\
\vfill {\bf Abstract}
\end{center}

The Lagrangian (action) formulation of the Chalker-Coddington
network model for plateau-plateau transitions in quantum Hall
effect is presented based on a model of fermions hopping on
Manhattan Lattice ($ML$). The dimensionless Landauer resistance is
considered and its average is calculated over the random $U(1)$
phases with constant distribution on the circle. The Lagrangian of
the resultant model on $ML$ is found and the corresponding
$R$-matrix is written down.
 It appeared, that this model is integrable, rising hope to investigate physics of
plateau- plateau transitions by the exact method of powerful
Algebraic Bethe Ansatz $(ABA)$.

\vfill
\rightline{January 2003}
\newpage
\pagestyle{plain}
1. The nature and the mechanism of plateau-plateau transitions in
the Integer Hall Effect ($IHE$) is still one of open problems in
the way of its complete understanding. In the article \cite{CC}
J.Chalker and P.Coddington ($CC$)  have introduced a network model
to deal with this problem and have studied it numerically. The
model is phenomenological and uses the Transfer Matrix formalism.
It based on the quasi-classical picture of noninteracting
electrons in two space dimensions moving along the boundaries of
the droplets(edge currents), which are formed by domains of the
constant perpendicular magnetic fields. This domains appear due to
disorder character of the magnetic field, presence of which is
crucial for the Quantum Hall Effect ($QHE$). Close to critical
point the droplets approaching each other and the quantum
tunnelling of edge electrons from one to other droplet becomes
essential. This phenomena causes the appearance of de-localized
state after disorder over random $U(1)$ phases are taken into
account.

The careful numerical investigations of the critical conductivity
and the correlation length index $\nu$ within the $CC$ network
model in a numerous articles \cite{CC, LWK, ChoF1, BH} have shown
excellent coincidence with the well established experimental value
$\nu \simeq 2.3 \pm 0.1 \;\;$( may be precisely 7/3) \cite{WEK,McEuen},
which essentially has motivated the considerable interest to the
model and stimulated it's further investigation up to our days
\cite{DHL1, DHL2, Kim, MT1, Zirn1,GRS1, AJS,KM1,ChoF2,MCh1,BC}.

The main problem in this direction off course is the problem of
formulation of the field theory, which corresponds to critical
limit of $CC$ model after quenched disorder is taken into account.
But already before the disorder one needs to find the equivalence
of $CC$ Transfer Matrix formulation on a some kind of lattice.

In the early stage of investigations of the $CC$ model D.H.Lee  \cite{DHL1}
mapped original
model to an antiferromagnetic spin chain.
Later Lee, together with Wang \cite{DHL2},  have extended the model to the
replica limit of the
associated Hubbard model.
 In the articles \cite{MT1} authors have developed further this ideas
 by mapping the
problem of localization to the problem of diagonalization of some
one-dimensional
non-Hermitian Hamiltonian
of interacting bosons and fermions.  But it is absolutely clear, that one
should consider this mappings as an alternative
description of he $CC$ model, rather than exact correspondence.

At the same time the technique of supersymmetry was used by
M.Zirnbauer to average over the disorder in $CC$ model in the
article \cite{Zirn1} , where he has reduced the problem  to the
supersymmetric $\sigma$-model and an antiferromagnetic
supersymmetric spin-chain.

In the articles \cite{KHAC}  the $U(1)$  $CC$ model was extended
to spin case to describe the spin $QH$ transitions in a system of
noninteracting quasi-particles in $2D$.  Authors of the articles
\cite{GRS1} have used a supersymmetry representation of such
models  to obtain a mapping onto the $2D$ classical bond
percolation transition and get some critical exponents and
universal ratios analytically. It is also necessary to mention
recent investigation of the chiral universality class of the
localization problem and corresponding modification of the
original network model \cite{BC}.

In the interesting article \cite{AJS} the authors, instead of mapping $CC$
model into some kind
of Hamiltonian,  are analyzing the hierarchy of network models by real space
renormalization
technique in order to understand the   nature of localization-delocalization
transition. The multifractal properties of the generalized $CC$ model was
investigated in
the articles \cite{KM1}.

Some variant of $CC$ network model was used in the articles
\cite{ChoF2, RL, GRL, MCh1}
in order to study numerically (and analytically) the random bond $2D$
Ising model  by mapping one
onto the other.

Analyzing carefully the mentioned upper articles, where authors have reduced
the original Transfer Matrix formulation of the $CC$ network model
(just for which the correlation length index was proved is coinciding with the
experimental value of $QH$ plateau transitions) to some kind of
supersymmetric spin chain problem one can find out that their exact
equivalence remains obscure.

In this article, developing the results obtained in our previous
works \cite{S, APSS, S1}, we give the exact action (Lagrangian)
formulation of the original $CC$ Transfer Matrix model as a field
theory on $2D$ lattice. First we show, that before the disorder
over $U(1)$ phases is taken into account the $CC$ model is
equivalent to some inhomogeneous modification of the $XX$ model in
the background  of $U(1)$ field. Basic element of this
construction is the fermionized version of the standard
$R$-operator of the $XX$ model, but the Transfer Matrix is the
staggered product of $\pi/2$-rotated $R$-operators. We prove, that
in one particle sector the matrix elements of the Transfer Matrix
of the defined model precisely reproduces the Transfer Matrix of
the $CC$ model. Then, by introducing the fermionic coherent
states, we write down the action of the $CC$ model on the $2D$
Manhattan Lattice ($ML$). Further,in order to  investigate the
presence of delocalized states, we consider the Landauer
resistance in the model and take into account the disorder  over
$U(1)$ phases.
 As it was argued in the articles \cite{ATAF},
the averaged Landauer resistance in quasi-one dimensional systems
defines the double of  localization length of the theory. We have
considered  homogenous distribution of random $U(1)$ phases over
the circle, calculated the average $\langle T \otimes
T^{{\dagger}} \rangle$($T$ is the Transfer Matrix of the $CC$
model) and found the $R$-matrix and the action of corresponding
model. The result is written in terms of two type (spin up and
spin down) Fermi fields. It appeared, that the $R$-matrix we have
found is fulfilling the Yang-Baxter Equations ($YBE$), which were
defined in the article \cite{APSS} for the models with staggered
disposition of $R$-matrices. This means that the resultant model
is integrable. It rises hopes, that by use of powerful technique
of Algebraic Bethe Ansatz ($ABA$) \cite{Bax, FT} one can
investigate and calculate the critical properties of the $QHE$
exactly.

2.The basic element of our construction is the
$R_{i,j}$-matrix(operator) of the Algebraic Bethe Ansatz
technique, which acts on a direct product of the linear spaces
$V_i$ and $V_j$ of the quantum states at the chain sites $i$ and
$j$ respectively.
\begin{eqnarray}
\label{RVV} \check R_{i,j}=V_i \otimes V_j \rightarrow
V^{\prime}_i \otimes V^{\prime}_j.
\end{eqnarray}
Let us attach the Fock spaces $V_j$ of scalar fermions $c_j^+, \;
c_j$ to each site of the chain and consider the operator forms of two
types of
$R$-matrices of the $XX$-model in  braid formalism
\begin{eqnarray}
\label{XX}
&&\check R^{\pm}_{2j,2j\pm1}=\nn\\
&=&a_{\pm}n_{2j}n_{2j\pm1} + a_{\pm}(1-n_{2j})(1-n_{2j\pm1})+
n_{2j}(1-n_{2j\pm1})\\
&+&(a_{\pm}a_{\pm}+b_{\pm}b_{\pm})n_{2j\pm1}(1-n_{2j})
+b_{\pm}c_{2j}^+c_{2j\pm1}+b_{\pm}c_{2j\pm1}^+c_{2j}\nn\\
&=&:e^{\left[b_{\pm}c_{2j\pm1}^+c_{2j} +b_{\pm}c_{2j}^+c_{2j\pm1}
+(a_{\pm}-1)c_{2j}^+c_{2j}+(1-a_{\pm})c_{2j\pm1}^+c_{2j\pm1}\right]}:
,\nn
\end{eqnarray}
corresponding to two type of acts of scattering in the $CC$ model,
as it is drown in Figure1.
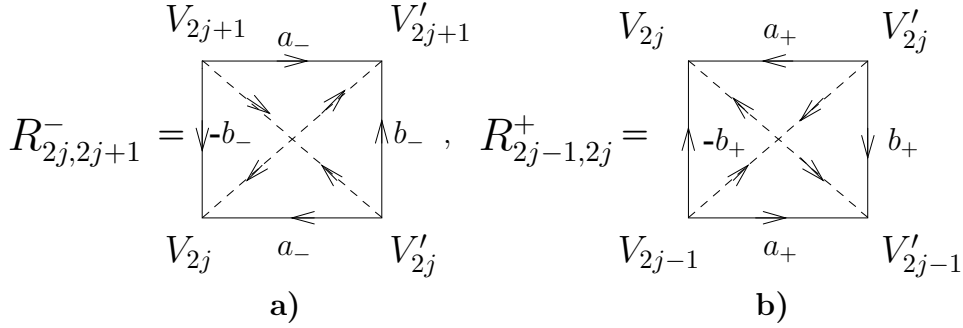
\begin{figure}[ht]
\begin{center}
\setlength{\unitlength}{0.00087489in}
\begingroup\makeatletter\ifx\SetFigFont\undefined%
\gdef\SetFigFont#1#2#3#4#5{%
  \reset@font\fontsize{#1}{#2pt}%
  \fontfamily{#3}\fontseries{#4}\fontshape{#5}%
  \selectfont}%
\fi\endgroup%
{\renewcommand{\dashlinestretch}{30}
\begin{picture}(5058,1909)(0,-10)
\path(720,1534)(720,634)(1800,634)
        (1800,1579)(720,1579)
\path(720,1579)(720,1534)
\path(3645,1534)(3645,634)(4725,634)
        (4725,1579)(3645,1579)
\path(3645,1579)(3645,1534) \dashline{60.000}(720,634)(1800,1579)
\dashline{60.000}(720,1579)(1800,634)
\dashline{60.000}(3645,634)(4725,1579)
\dashline{60.000}(3645,1579)(4725,634) \path(1395,634)(1260,634)
\path(1395.000,671.500)(1260.000,634.000)(1395.000,596.500)
\path(720,1219)(720,1039)
\path(682.500,1174.000)(720.000,1039.000)(757.500,1174.000)
\path(1170,1579)(1350,1579)
\path(1215.000,1541.500)(1350.000,1579.000)(1215.000,1616.500)
\path(1800,1084)(1800,1219)
\path(1837.500,1084.000)(1800.000,1219.000)(1762.500,1084.000)
\path(4050,634)(4230,634)
\path(4095.000,596.500)(4230.000,634.000)(4095.000,671.500)
\path(3645,1039)(3645,1174)
\path(3682.500,1039.000)(3645.000,1174.000)(3607.500,1039.000)
\path(4275,1579)(4095,1579)
\path(4230.000,1616.500)(4095.000,1579.000)(4230.000,1541.500)
\path(4725,1174)(4725,994)
\path(4687.500,1129.000)(4725.000,994.000)(4762.500,1129.000)
\path(1080,949)(990,859)
\path(1058.943,980.976)(990.000,859.000)(1111.976,927.943)
\path(1440,1264)(1575,1399)
\path(1506.057,1277.024)(1575.000,1399.000)(1453.024,1330.057)
\path(1530,859)(1440,949)
\path(1561.976,880.057)(1440.000,949.000)(1508.943,827.024)
\path(990,1354)(1125,1264)
\path(991.872,1307.683)(1125.000,1264.000)(1033.474,1370.086)
\path(3915,859)(4005,949)
\path(3936.057,827.024)(4005.000,949.000)(3883.024,880.057)
\path(4455,1354)(4320,1219)
\path(4388.943,1340.976)(4320.000,1219.000)(4441.976,1287.943)
\path(4320,994)(4455,859)
\path(4333.024,927.943)(4455.000,859.000)(4386.057,980.976)
\path(4005,1264)(3915,1354)
\path(4036.976,1285.057)(3915.000,1354.000)(3983.943,1232.024)
\put(-450,1039){\makebox(0,0)[lb]{\smash{{{\SetFigFont{16}{20.4}{\rmdefault}
{\bfdefault} {\updefault}$R^-_{2j,2j+1}$}}}}}
\put(2385,1039){\makebox(0,0)[lb]{\smash{{{\SetFigFont{16}{20.4}{\rmdefault}
{\bfdefault} {\updefault}$R^+_{2j-1,2j}$}}}}}
\put(495,1759){\makebox(0,0)[lb]{\smash{{{\SetFigFont{14}{16.8}{\rmdefault}
{\bfdefault} {\updefault}$V_{2j+1}$}}}}}
\put(495,364){\makebox(0,0)[lb]{\smash{{{\SetFigFont{14}{16.8}{\rmdefault}
{\bfdefault} {\updefault}$V_{2j}$}}}}}
\put(1845,1759){\makebox(0,0)[lb]{\smash{{{\SetFigFont{14}{16.8}{\rmdefault}
{\bfdefault} {\updefault}$V^{\prime}_{2j+1}$}}}}}
\put(1845,364){\makebox(0,0)[lb]{\smash{{{\SetFigFont{14}{16.8}{\rmdefault}
{\bfdefault} {\updefault}$V^{\prime}_{2j}$}}}}}
\put(3220,364){\makebox(0,0)[lb]{\smash{{{\SetFigFont{14}{16.8}{\rmdefault}
{\bfdefault} {\updefault}$V_{2j-1}$}}}}}
\put(4805,364){\makebox(0,0)[lb]{\smash{{{\SetFigFont{14}{16.8}{\rmdefault}
{\bfdefault} {\updefault}$V^{\prime}_{2j-1}$}}}}}
\put(3220,1714){\makebox(0,0)[lb]{\smash{{{\SetFigFont{14}{16.8}{\rmdefault}
{\bfdefault} {\updefault}$V_{2j}$}}}}}
\put(4805,1714){\makebox(0,0)[lb]{\smash{{{\SetFigFont{14}{16.8}{\rmdefault}
{\bfdefault} {\updefault}$V^{\prime}_{2j}$}}}}}
\put(3240,1084){\makebox(0,0)[lb]{\smash{{{\SetFigFont{12}{16.8}{\rmdefault}
{\bfdefault} {\updefault}=}}}}}
\put(765,1084){\makebox(0,0)[lb]{\smash{{{\SetFigFont{12}{14.4}{\rmdefault}
{\bfdefault} {\updefault}-$b_{-}$}}}}}
\put(3735,1039){\makebox(0,0)[lb]{\smash{{{\SetFigFont{12}{14.4}{\rmdefault}
{\bfdefault} {\updefault}-$b_{+}$}}}}}
\put(4845,1039){\makebox(0,0)[lb]{\smash{{{\SetFigFont{12}{14.4}{\rmdefault}
{\bfdefault} {\updefault}$b_{+}$}}}}}
\put(1170,1669){\makebox(0,0)[lb]{\smash{{{\SetFigFont{12}{14.4}{\rmdefault}
{\bfdefault} {\updefault}$a_{-}$}}}}}
\put(1170,409){\makebox(0,0)[lb]{\smash{{{\SetFigFont{12}{14.4}{\rmdefault}
{\bfdefault} {\updefault}$a_{-}$}}}}}
\put(4095,409){\makebox(0,0)[lb]{\smash{{{\SetFigFont{12}{14.4}{\rmdefault}
{\bfdefault} {\updefault}$a_{+}$}}}}}
\put(4095,1714){\makebox(0,0)[lb]{\smash{{{\SetFigFont{12}{14.4}{\rmdefault}
{\bfdefault} {\updefault}$a_{+}$}}}}}
\put(4050,49){\makebox(0,0)[lb]{\smash{{{\SetFigFont{13}{16.8}{\rmdefault}
{\bfdefault} {\updefault}b)}}}}}
\put(1125,49){\makebox(0,0)[lb]{\smash{{{\SetFigFont{13}{16.8}{\rmdefault}
{\bfdefault} {\updefault}a)}}}}}
\put(1870,1084){\makebox(0,0)[lb]{\smash{{{\SetFigFont{12}{14.4}{\rmdefault}
{\bfdefault} {\updefault}$b_{-}\;\;,$}}}}}
\put(450,1084){\makebox(0,0)[lb]{\smash{{{\SetFigFont{12}{16.8}{\rmdefault}
{\bfdefault} {\updefault}=}}}}}
\end{picture}}
\vspace{-0.7cm}
\end{center}
\caption{The two type $R$-matrices in $CC$ model}
\end{figure}

In the expression (\ref{XX})  the symbol $: \;\;:$  means normal
ordering of fermionic operators in the even sites and anti-normal
(hole) ordering in the odd sites. The convenience of this choose
will be clear later. The dot lines in the picture represents
standard view of the $R$-matrices, while solid lines are
convenient in a language of fermions with hopping parameters
$a_{\pm}$ and $b_{\pm}$. Each of the $ R^{\pm}$ operators are
nothing, but the fermionized versions of $R$-matrices of the
ordinary $XX$ models
\begin{eqnarray}
\label{XXR}
\check{R}_{\pm} = \left(\begin{array}{llll}
a_{\pm}&0&0&0\\
0&1&b_{\pm}&0\\
0&b_{\pm}&(a_{\pm}a_{\pm}+b_{\pm}b_{\pm})&0\\
0&0&0&a_{\pm}

\end{array}\right),
\end{eqnarray}
which can be found by Jordan-Wigner transformation \cite{W},  or by
the alternative technique, developed in \cite{AKMS}.

Let us consider
now Monodromy matrices $M_1$ and $M_2$ as a following products of
$R$-matrices
\begin{equation}
\label{MRR} M_1=\prod^N_{j=1} \check{R}^+_{2j-1,2j},\qquad
M_2=\prod^N_{j=1} \check{R}^-_{2j,2j+1},
\end{equation}
which are corresponding to neighbor columns  of the scatterings in
the Figure 2. $M_{1,2}$ are discrete time  evolution operators of
the quantum states of the chain  $\mid \Psi(t)\rangle \in \prod_j
V_j$ on one step $\mid \Psi(t+1)\rangle= M \mid \Psi(t)\rangle$ .
They are acting consequently and, therefore, the translational
invariance in a time and chain directions exists only for even
lattice spacing translations.
\begin{figure}[ht]
\begin{center}
\setlength{\unitlength}{0.00087489in}
\begingroup\makeatletter\ifx\SetFigFont\undefined%
\gdef\SetFigFont#1#2#3#4#5{%
  \reset@font\fontsize{#1}{#2pt}%
  \fontfamily{#3}\fontseries{#4}\fontshape{#5}%
  \selectfont}%
\fi\endgroup%
{\renewcommand{\dashlinestretch}{30}
\begin{picture}(6717,4797)(0,-10)
\drawline(405,4320)(405,4320)
\drawline(855,4320)(855,4320)
\path(405,4320)(6705,4320)
\path(405,3420)(6705,3420)
\path(405,2520)(6705,2520)
\path(405,1620)(6705,1620)
\path(405,720)(6705,720)
\path(855,4770)(855,270)
\path(1755,4770)(1755,270)
\path(2655,4770)(2655,270)
\path(3555,4770)(3555,270)
\path(4455,4770)(4455,270)
\path(5355,4770)(5355,270)
\path(6255,4770)(6255,270)
\drawline(1305,4320)(1305,4320)
\drawline(1305,4320)(1305,4320)
\path(855,4320)(1305,4320)
\path(1185.000,4290.000)(1305.000,4320.000)(1185.000,4350.000)
\path(1755,4320)(2205,4320)
\path(2085.000,4290.000)(2205.000,4320.000)(2085.000,4350.000)
\path(2655,4320)(3105,4320)
\path(2985.000,4290.000)(3105.000,4320.000)(2985.000,4350.000)
\path(3555,4320)(4005,4320)
\path(3885.000,4290.000)(4005.000,4320.000)(3885.000,4350.000)
\path(4455,4320)(4905,4320)
\path(4785.000,4290.000)(4905.000,4320.000)(4785.000,4350.000)
\path(5355,4320)(5805,4320)
\path(5685.000,4290.000)(5805.000,4320.000)(5685.000,4350.000)
\path(6255,4320)(6705,4320)
\path(6585.000,4290.000)(6705.000,4320.000)(6585.000,4350.000)
\path(855,2520)(1305,2520)
\path(1185.000,2490.000)(1305.000,2520.000)(1185.000,2550.000)
\path(1755,2520)(2205,2520)
\path(2085.000,2490.000)(2205.000,2520.000)(2085.000,2550.000)
\path(2655,2520)(3105,2520)
\path(2985.000,2490.000)(3105.000,2520.000)(2985.000,2550.000)
\path(3555,2520)(4005,2520)
\path(3885.000,2490.000)(4005.000,2520.000)(3885.000,2550.000)
\path(4455,2520)(4905,2520)
\path(4785.000,2490.000)(4905.000,2520.000)(4785.000,2550.000)
\path(5355,2520)(5805,2520)
\path(5685.000,2490.000)(5805.000,2520.000)(5685.000,2550.000)
\path(855,720)(1305,720)
\path(1185.000,690.000)(1305.000,720.000)(1185.000,750.000)
\path(1755,720)(2205,720)
\path(2085.000,690.000)(2205.000,720.000)(2085.000,750.000)
\path(2655,720)(3105,720)
\path(2985.000,690.000)(3105.000,720.000)(2985.000,750.000)
\path(3555,720)(4005,720)
\path(3885.000,690.000)(4005.000,720.000)(3885.000,750.000)
\path(4455,720)(4905,720)
\path(4785.000,690.000)(4905.000,720.000)(4785.000,750.000)
\drawline(5580,720)(5580,720)
\path(855,3420)(855,3870)
\path(885.000,3750.000)(855.000,3870.000)(825.000,3750.000)
\path(855,2520)(855,2970)
\path(885.000,2850.000)(855.000,2970.000)(825.000,2850.000)
\path(855,1620)(855,2070)
\path(885.000,1950.000)(855.000,2070.000)(825.000,1950.000)
\path(855,720)(855,1170)
\path(885.000,1050.000)(855.000,1170.000)(825.000,1050.000)
\path(855,4320)(855,4770)
\path(885.000,4650.000)(855.000,4770.000)(825.000,4650.000)
\path(1755,4320)(1755,3870)
\path(1725.000,3990.000)(1755.000,3870.000)(1785.000,3990.000)
\path(1755,3420)(1755,2970)
\path(1725.000,3090.000)(1755.000,2970.000)(1785.000,3090.000)
\path(1755,2520)(1755,2070)
\path(1725.000,2190.000)(1755.000,2070.000)(1785.000,2190.000)
\path(1755,1620)(1755,1170)
\path(1725.000,1290.000)(1755.000,1170.000)(1785.000,1290.000)
\path(1755,720)(1755,270)
\path(1725.000,390.000)(1755.000,270.000)(1785.000,390.000)
\path(2655,3420)(2655,3870)
\path(2685.000,3750.000)(2655.000,3870.000)(2625.000,3750.000)
\path(2655,4320)(2655,4770)
\path(2685.000,4650.000)(2655.000,4770.000)(2625.000,4650.000)
\path(2655,2520)(2655,2970)
\path(2685.000,2850.000)(2655.000,2970.000)(2625.000,2850.000)
\path(2655,1620)(2655,2070)
\path(2685.000,1950.000)(2655.000,2070.000)(2625.000,1950.000)
\path(2655,720)(2655,1170)
\path(2685.000,1050.000)(2655.000,1170.000)(2625.000,1050.000)
\path(3555,4320)(3555,3870)
\path(3525.000,3990.000)(3555.000,3870.000)(3585.000,3990.000)
\path(3555,3420)(3555,2970)
\path(3525.000,3090.000)(3555.000,2970.000)(3585.000,3090.000)
\path(3555,2520)(3555,2070)
\path(3525.000,2190.000)(3555.000,2070.000)(3585.000,2190.000)
\path(3555,1620)(3555,1170)
\path(3525.000,1290.000)(3555.000,1170.000)(3585.000,1290.000)
\path(3555,720)(3555,270)
\path(3525.000,390.000)(3555.000,270.000)(3585.000,390.000)
\path(4455,720)(4455,1170)
\path(4485.000,1050.000)(4455.000,1170.000)(4425.000,1050.000)
\path(4455,1620)(4455,2070)
\path(4485.000,1950.000)(4455.000,2070.000)(4425.000,1950.000)
\path(4455,2520)(4455,2970)
\path(4485.000,2850.000)(4455.000,2970.000)(4425.000,2850.000)
\path(4455,3420)(4455,3870)
\path(4485.000,3750.000)(4455.000,3870.000)(4425.000,3750.000)
\path(4455,4320)(4455,4770)
\path(4485.000,4650.000)(4455.000,4770.000)(4425.000,4650.000)
\path(5355,4320)(5355,3870)
\path(5325.000,3990.000)(5355.000,3870.000)(5385.000,3990.000)
\path(5355,3420)(5355,2970)
\path(5325.000,3090.000)(5355.000,2970.000)(5385.000,3090.000)
\path(5355,2520)(5355,2070)
\path(5325.000,2190.000)(5355.000,2070.000)(5385.000,2190.000)
\path(5355,1620)(5355,1170)
\path(5325.000,1290.000)(5355.000,1170.000)(5385.000,1290.000)
\path(5355,720)(5355,270)
\path(5325.000,390.000)(5355.000,270.000)(5385.000,390.000)
\path(6255,3420)(6255,3870)
\path(6285.000,3750.000)(6255.000,3870.000)(6225.000,3750.000)
\path(6255,4320)(6255,4770)
\path(6285.000,4650.000)(6255.000,4770.000)(6225.000,4650.000)
\path(6255,2520)(6255,2970)
\path(6285.000,2850.000)(6255.000,2970.000)(6225.000,2850.000)
\path(6255,1620)(6255,2070)
\path(6285.000,1950.000)(6255.000,2070.000)(6225.000,1950.000)
\path(6255,720)(6255,1170)
\path(6285.000,1050.000)(6255.000,1170.000)(6225.000,1050.000)
\path(6255,270)(6255,720)
\path(6285.000,600.000)(6255.000,720.000)(6225.000,600.000)
\path(1755,3420)(1305,3420)
\path(1425.000,3450.000)(1305.000,3420.000)(1425.000,3390.000)
\path(855,3420)(405,3420)
\path(525.000,3450.000)(405.000,3420.000)(525.000,3390.000)
\path(2655,3420)(2205,3420)
\path(2325.000,3450.000)(2205.000,3420.000)(2325.000,3390.000)
\path(3555,3420)(3105,3420)
\path(3225.000,3450.000)(3105.000,3420.000)(3225.000,3390.000)
\path(4455,3420)(4005,3420)
\path(4125.000,3450.000)(4005.000,3420.000)(4125.000,3390.000)
\path(5355,3420)(4905,3420)
\path(5025.000,3450.000)(4905.000,3420.000)(5025.000,3390.000)
\path(6255,3420)(5805,3420)
\path(5925.000,3450.000)(5805.000,3420.000)(5925.000,3390.000)
\path(855,1620)(405,1620)
\path(525.000,1650.000)(405.000,1620.000)(525.000,1590.000)
\path(1755,1620)(1305,1620)
\path(1425.000,1650.000)(1305.000,1620.000)(1425.000,1590.000)
\path(2655,1620)(2205,1620)
\path(2325.000,1650.000)(2205.000,1620.000)(2325.000,1590.000)
\path(3555,1620)(3105,1620)
\path(3225.000,1650.000)(3105.000,1620.000)(3225.000,1590.000)
\path(4455,1620)(4005,1620)
\path(4125.000,1650.000)(4005.000,1620.000)(4125.000,1590.000)
\path(5355,1620)(4905,1620)
\path(5025.000,1650.000)(4905.000,1620.000)(5025.000,1590.000)
\path(6255,1620)(5805,1620)
\path(5925.000,1650.000)(5805.000,1620.000)(5925.000,1590.000)
\path(6255,720)(6705,720)
\path(6585.000,690.000)(6705.000,720.000)(6585.000,750.000)
\path(6255,2520)(6705,2520)
\path(6585.000,2490.000)(6705.000,2520.000)(6585.000,2550.000)
\dashline{60.000}(855,2520)(1305,2970)
\path(1241.360,2863.934)(1305.000,2970.000)(1198.934,2906.360)
\dashline{60.000}(1755,3420)(1305,2970)
\path(1368.640,3076.066)(1305.000,2970.000)(1411.066,3033.640)
\dashline{60.000}(1305,2970)(855,3420)
\path(961.066,3356.360)(855.000,3420.000)(918.640,3313.934)
\dashline{60.000}(1305,2970)(1755,2520)
\path(1648.934,2583.640)(1755.000,2520.000)(1691.360,2626.066)
\dashline{60.000}(1755,2520)(2205,2070)
\path(2098.934,2133.640)(2205.000,2070.000)(2141.360,2176.066)
\dashline{60.000}(2655,1620)(2205,2070)
\path(2311.066,2006.360)(2205.000,2070.000)(2268.640,1963.934)
\dashline{60.000}(3105,1170)(2655,1620)
\path(2761.066,1556.360)(2655.000,1620.000)(2718.640,1513.934)
\dashline{60.000}(3105,1170)(3555,720)
\path(3448.934,783.640)(3555.000,720.000)(3491.360,826.066)
\dashline{60.000}(1305,1170)(855,1620)
\path(961.066,1556.360)(855.000,1620.000)(918.640,1513.934)
\dashline{60.000}(1305,1170)(1755,720)
\path(1648.934,783.640)(1755.000,720.000)(1691.360,826.066)
\dashline{60.000}(855,720)(1305,1170)
\path(1241.360,1063.934)(1305.000,1170.000)(1198.934,1106.360)
\dashline{60.000}(1755,1620)(1305,1170)
\path(1368.640,1276.066)(1305.000,1170.000)(1411.066,1233.640)
\dashline{60.000}(2205,2070)(1755,1620)
\path(1818.640,1726.066)(1755.000,1620.000)(1861.066,1683.640)
\dashline{60.000}(2205,2070)(2655,2520)
\path(2591.360,2413.934)(2655.000,2520.000)(2548.934,2456.360)
\dashline{60.000}(2655,2520)(3105,2970)
\path(3041.360,2863.934)(3105.000,2970.000)(2998.934,2906.360)
\dashline{60.000}(3555,3420)(3105,2970)
\path(3168.640,3076.066)(3105.000,2970.000)(3211.066,3033.640)
\dashline{60.000}(4005,3870)(3555,3420)
\path(3618.640,3526.066)(3555.000,3420.000)(3661.066,3483.640)
\dashline{60.000}(4005,3870)(4455,4320)
\path(4391.360,4213.934)(4455.000,4320.000)(4348.934,4256.360)
\dashline{60.000}(1755,4320)(2205,3870)
\path(2098.934,3933.640)(2205.000,3870.000)(2141.360,3976.066)
\dashline{60.000}(2205,3870)(1755,3420)
\path(1818.640,3526.066)(1755.000,3420.000)(1861.066,3483.640)
\dashline{60.000}(2205,3870)(2655,4320)
\path(2591.360,4213.934)(2655.000,4320.000)(2548.934,4256.360)
\dashline{60.000}(2655,3420)(2205,3870)
\path(2311.066,3806.360)(2205.000,3870.000)(2268.640,3763.934)
\dashline{60.000}(3105,2970)(2655,3420)
\path(2761.066,3356.360)(2655.000,3420.000)(2718.640,3313.934)
\dashline{60.000}(3105,2970)(3555,2520)
\path(3448.934,2583.640)(3555.000,2520.000)(3491.360,2626.066)
\dashline{60.000}(3555,2520)(4005,2070)
\path(3898.934,2133.640)(4005.000,2070.000)(3941.360,2176.066)
\dashline{60.000}(4455,1620)(4005,2070)
\path(4111.066,2006.360)(4005.000,2070.000)(4068.640,1963.934)
\dashline{60.000}(4905,1170)(4455,1620)
\path(4561.066,1556.360)(4455.000,1620.000)(4518.640,1513.934)
\dashline{60.000}(4905,1170)(5355,720)
\path(5248.934,783.640)(5355.000,720.000)(5291.360,826.066)
\dashline{60.000}(3555,4320)(4005,3870)
\path(3898.934,3933.640)(4005.000,3870.000)(3941.360,3976.066)
\dashline{60.000}(4455,3420)(4005,3870)
\path(4111.066,3806.360)(4005.000,3870.000)(4068.640,3763.934)
\dashline{60.000}(4905,2970)(4455,3420)
\path(4561.066,3356.360)(4455.000,3420.000)(4518.640,3313.934)
\dashline{60.000}(4905,2970)(5355,2520)
\path(5248.934,2583.640)(5355.000,2520.000)(5291.360,2626.066)
\dashline{60.000}(5355,2520)(5805,2070)
\path(5698.934,2133.640)(5805.000,2070.000)(5741.360,2176.066)
\dashline{60.000}(6255,1620)(5805,2070)
\path(5911.066,2006.360)(5805.000,2070.000)(5868.640,1963.934)
\dashline{60.000}(2655,720)(3105,1170)
\path(3041.360,1063.934)(3105.000,1170.000)(2998.934,1106.360)
\dashline{60.000}(3555,1620)(3105,1170)
\path(3168.640,1276.066)(3105.000,1170.000)(3211.066,1233.640)
\dashline{60.000}(4005,2070)(3555,1620)
\path(3618.640,1726.066)(3555.000,1620.000)(3661.066,1683.640)
\dashline{60.000}(4005,2070)(4455,2520)
\path(4391.360,2413.934)(4455.000,2520.000)(4348.934,2456.360)
\dashline{60.000}(4455,2520)(4905,2970)
\path(4841.360,2863.934)(4905.000,2970.000)(4798.934,2906.360)
\dashline{60.000}(5355,3420)(4905,2970)
\path(4968.640,3076.066)(4905.000,2970.000)(5011.066,3033.640)
\dashline{60.000}(5805,3870)(5355,3420)
\path(5418.640,3526.066)(5355.000,3420.000)(5461.066,3483.640)
\dashline{60.000}(5805,3870)(6255,4320)
\path(6191.360,4213.934)(6255.000,4320.000)(6148.934,4256.360)
\dashline{60.000}(5355,4320)(5805,3870)
\path(5698.934,3933.640)(5805.000,3870.000)(5741.360,3976.066)
\dashline{60.000}(6255,3420)(5805,3870)
\path(5911.066,3806.360)(5805.000,3870.000)(5868.640,3763.934)
\dashline{60.000}(5805,2070)(5355,1620)
\path(5418.640,1726.066)(5355.000,1620.000)(5461.066,1683.640)
\dashline{60.000}(5355,1620)(4905,1170)
\path(4968.640,1276.066)(4905.000,1170.000)(5011.066,1233.640)
\dashline{60.000}(4455,720)(4905,1170)
\path(4841.360,1063.934)(4905.000,1170.000)(4798.934,1106.360)
\dashline{60.000}(5805,2070)(6255,2520)
\path(6191.360,2413.934)(6255.000,2520.000)(6148.934,2456.360)
\dashline{60.000}(5355,720)(5805,720)
\path(5685.000,690.000)(5805.000,720.000)(5685.000,750.000)
\put(0,720){\makebox(0,0)[lb]{\smash{{{\SetFigFont{12}{14.4}{\rmdefault}{\mddefault}
{\updefault}2j-1}}}}}
\put(0,1665){\makebox(0,0)[lb]{\smash{{{\SetFigFont{12}{14.4}{\rmdefault}{\mddefault}
{\updefault}2j}}}}}
\put(0,2565){\makebox(0,0)[lb]{\smash{{{\SetFigFont{12}{14.4}{\rmdefault}{\mddefault}
{\updefault}2j+1}}}}}
\put(0,3465){\makebox(0,0)[lb]{\smash{{{\SetFigFont{12}{14.4}{\rmdefault}{\mddefault}
{\updefault}2j+2}}}}}
\put(1755,0){\makebox(0,0)[lb]{\smash{{{\SetFigFont{12}{14.4}{\rmdefault}{\mddefault}
{\updefault}t+1}}}}}
\put(2655,45){\makebox(0,0)[lb]{\smash{{{\SetFigFont{12}{14.4}{\rmdefault}{\mddefault}
{\updefault}t+2}}}}}
\put(3555,45){\makebox(0,0)[lb]{\smash{{{\SetFigFont{12}{14.4}{\rmdefault}{\mddefault}
{\updefault}t+3}}}}}
\put(855,0){\makebox(0,0)[lb]{\smash{{{\SetFigFont{12}{14.4}{\rmdefault}{\mddefault}
{\updefault}t}}}}}
\put(2205,45){\makebox(0,0)[lb]{\smash{{{\SetFigFont{14}{16.8}{\rmdefault}{\mddefault}
{\updefault}$M_2$}}}}}
\put(3105,45){\makebox(0,0)[lb]{\smash{{{\SetFigFont{14}{16.8}{\rmdefault}{\mddefault}
{\updefault}$M_1$}}}}}
\put(4005,45){\makebox(0,0)[lb]{\smash{{{\SetFigFont{14}{16.8}{\rmdefault}{\mddefault}
{\updefault}$M_2$}}}}}
\put(4905,45){\makebox(0,0)[lb]{\smash{{{\SetFigFont{14}{16.8}{\rmdefault}{\mddefault}
{\updefault}$M_1$}}}}}
\put(5805,45){\makebox(0,0)[lb]{\smash{{{\SetFigFont{14}{16.8}{\rmdefault}{\mddefault}
{\updefault}$M_2$}}}}}
\put(1305,45){\makebox(0,0)[lb]{\smash{{{\SetFigFont{14}{16.8}{\rmdefault}{\mddefault}
{\updefault}$M_1$}}}}}
\put(540,1080){\makebox(0,0)[lb]{\smash{{{\SetFigFont{12}{14.4}{\rmdefault}
{\mddefault}
{\updefault}-$b_+$}}}}}
\put(1800,1080){\makebox(0,0)[lb]{\smash{{{\SetFigFont{12}{14.4}{\rmdefault}
{\mddefault}
{\updefault}$b_+$}}}}}
\put(1170,1710){\makebox(0,0)[lb]{\smash{{{\SetFigFont{12}{14.4}{\rmdefault}
{\mddefault}
{\updefault}$a_+$}}}}}
\put(1170,495){\makebox(0,0)[lb]{\smash{{{\SetFigFont{12}{14.4}{\rmdefault}
{\mddefault}
{\updefault}$a_+$}}}}}
\put(2115,1395){\makebox(0,0)[lb]{\smash{{{\SetFigFont{12}{14.4}{\rmdefault}
{\mddefault}
{\updefault}$a_-$}}}}}
\put(2115,2610){\makebox(0,0)[lb]{\smash{{{\SetFigFont{12}{14.4}{\rmdefault}
{\mddefault}
{\updefault}$a_-$}}}}}
\put(1485,2070){\makebox(0,0)[lb]{\smash{{{\SetFigFont{12}{14.4}{\rmdefault}
{\mddefault}
{\updefault}-$b_-$}}}}}
\put(2745,2070){\makebox(0,0)[lb]{\smash{{{\SetFigFont{12}{14.4}{\rmdefault}
{\mddefault}
{\updefault}$b_-$}}}}}
\put(585,2970){\makebox(0,0)[lb]{\smash{{{\SetFigFont{12}{14.4}{\rmdefault}
{\mddefault}
{\updefault}-$b_+$}}}}}
\put(1845,2970){\makebox(0,0)[lb]{\smash{{{\SetFigFont{12}{14.4}{\rmdefault}
{\mddefault}
{\updefault}$b_+$}}}}}
\put(1170,2340){\makebox(0,0)[lb]{\smash{{{\SetFigFont{12}{14.4}{\rmdefault}
{\mddefault}
{\updefault}$a_+$}}}}}
\put(1170,3465){\makebox(0,0)[lb]{\smash{{{\SetFigFont{12}{14.4}{\rmdefault}
{\mddefault}
{\updefault}$a_+$}}}}}
\put(1485,3870){\makebox(0,0)[lb]{\smash{{{\SetFigFont{12}{14.4}{\rmdefault}
{\mddefault}
{\updefault}-$b_-$}}}}}
\put(2745,3870){\makebox(0,0)[lb]{\smash{{{\SetFigFont{12}{14.4}{\rmdefault}
{\mddefault}
{\updefault}$b_-$}}}}}
\put(2070,3240){\makebox(0,0)[lb]{\smash{{{\SetFigFont{12}{14.4}{\rmdefault}
{\mddefault}
{\updefault}$a_-$}}}}}
\put(2070,4410){\makebox(0,0)[lb]{\smash{{{\SetFigFont{12}{14.4}{\rmdefault}
{\mddefault}
{\updefault}$a_-$}}}}}
\end{picture}}
\vspace{-0.7cm}
\end{center}
\caption{The dot lines represent the $CC$ model. The solid lines
represents the Manhattan Lattice, where equivalent formulation of
$CC$ can be given.}
\end{figure}
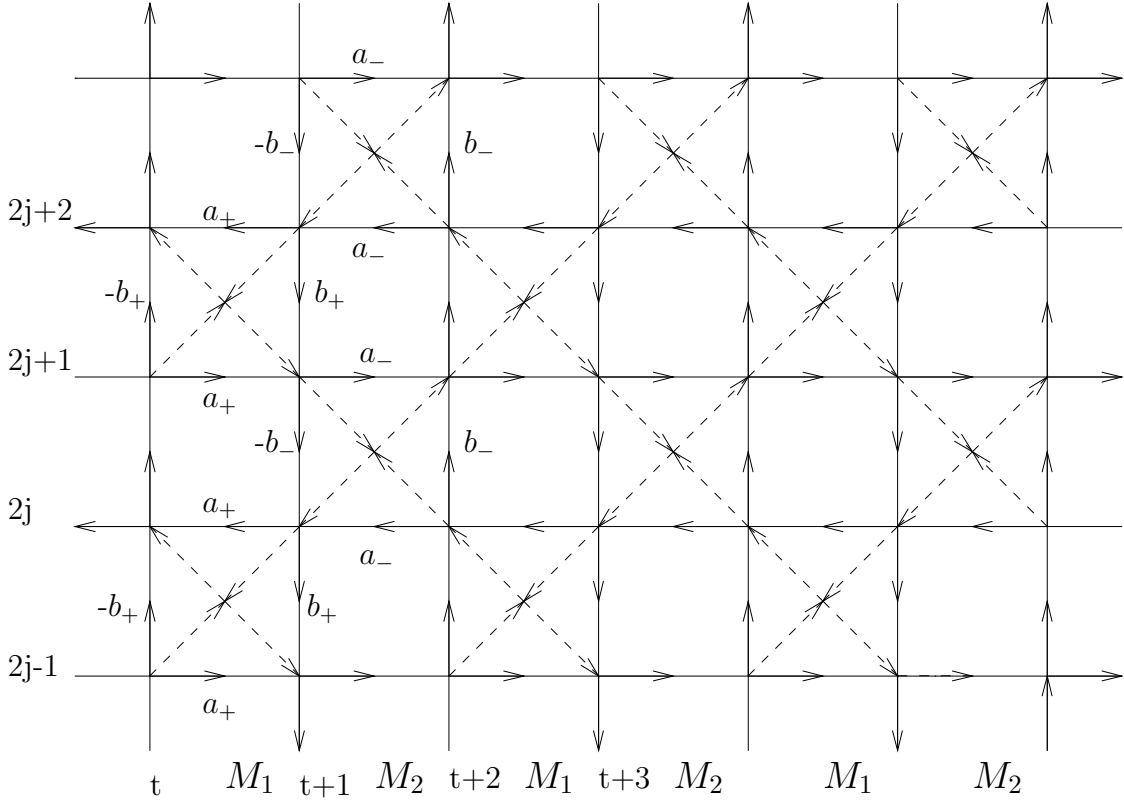

The dot lines in the Figure 2. represents standard $CC$ model,
while solid lines define so called Manhattan Lattice ($ML$) due to
the disposition of arrows on links. We have assigned the links of
$ML$ with  corresponding hopping parameter.

Let us consider now the one-particle sector of the Fock space of the chain
\begin{eqnarray}
\label{OP}
\mid i\rangle= \mid 0,0..1_i,..0\rangle=c^+_i \mid 0,0..0,..0\rangle,
\qquad\qquad i=1,...2N
\end{eqnarray}
and calculate the matrix elements of the operators $M_1$ and $M_2$
in this basis. After the parametrization of the hopping parameters
as
\begin{eqnarray}
\label{hop}
a_{-}= 1/\cosh\theta^{\prime},\qquad b_{-}= 1/\tanh\theta^{\prime},\qquad
a_{+}=a= 1/\cosh\theta,\qquad
b_{+}=b= 1/\tanh\theta
\end{eqnarray}
and unessential rescaling of $M_1$ and $M_2$ by the factors
$a^N_{+}$ and $a^N_{-}$ respectively, one easily can recover the
$2N\; X\; 2N$ Transfer matrices $B$ and $C$, introduced in the
article \cite{CC} by J.Chalker and P.Coddington. In order to have
$2D$ rotational invariance they have fixed
$a_{+}=b_{-}=a,\;b_{+}=a_{-}=b$, which gives
$\sinh\theta=\sinh\theta^{\prime}$ {\footnote {It is necessary now
to make some remark. The rotational invariance indeed  demands
this relations between the hopping parameters, as it can be seen
in the Figure 2. by rotating  the $R$-matrices on $\pi/2$, but the
disposition of signs will be different. However, since it does not
changes the result after the calculation of the disorder we left
it as in \cite{CC}}}.

The random $U(1)$-phase factors ($A$ and $C$ matrices in
\cite{CC})  can be introduced via operators $U_{\{\alpha\}}=\exp[i
\sum^{2N}_{j=1}\alpha_j n_j]=U_{\alpha_{even}}U_{\alpha_{odd}}$.
The Transfer Matrix for the slice will look now as $M=M_1 U_1 M_2
U_2$ and it defines the $N$-step evolution operator by $T = (M_1
U_1 M_2 U_2)^N =
(U_{\alpha_{even}}M_1U_{\alpha_{odd}}U_{\alpha_{even}}M_2U_{\alpha_{odd}})^N$.
The phase factors in the operator
$U_{\alpha_{even}}M_{1,2}U_{\alpha_{odd}}$ appear near the
$a_{\pm}, b_{\pm}$ parameters of the expression (\ref{XX}) for the
$R$-operators, and we obtain
\begin{eqnarray}
\label{dXX}
&&\bar{R}^{\pm}_{2j,2j\pm1}=e^{i \alpha n_{2j}}\check R^{\pm}_{2j,2j\pm1}
e^{i \alpha^{\prime} n_{2j\pm1}} \\
&=&:e^{\left[e^{i \alpha}b_{\pm}c_{2j\pm1}^+c_{2j} +e^{i \alpha^{\prime}}
b_{\pm}c_{2j}^+c_{2j\pm1}
+(a_{\pm}e^{i \alpha}-1)c_{2j}^+c_{2j}+
(1-a_{\pm}e^{i \alpha^{\prime}})c_{2j\pm1}^+c_{2j\pm1}\right]}:\nn
\end{eqnarray}
As we see from this formula, the phase factors are the same on the
exiting from the point links of the graphical representation of
the $R $-matrix (see Fig.1).

Now we would like to present a field theory formulation of the
$CC$ model on $ML$ and define an action for it. In Lagrangian
formulation the calculation of phase disorder for the Landauer
resistance will be straightforward. Let us introduce fermionic
coherent states according to articles \cite{F} and pass to
fermionic Transfer Matrix as it is done in \cite{S,S1}.
\begin{eqnarray}
\label{CS1}
|\psi_{2j}\rangle = e^{\psi_{2j}c^+_{2j}}|0\rangle ,\qquad
\langle\bar{\psi}_{2j}| = \langle0| e^{c_{2j}\bar{\psi}_{2j}}
\end{eqnarray}
for the even sites of the chain and
\begin{eqnarray}
\label{CS2}
|\bar{\psi}_{2j+1}\rangle = (c^+_{2j+1}-\bar{\psi}_{2j+1})
|0\rangle ,\qquad
\langle\psi_{2j+1}| = \langle 0|(c_{2j+1}+\psi_{2j+1})
\end{eqnarray}
for the odd sites.

This states are designed as an eigenstates of
creation-annihilation operators of fermions $c_j^+,\;\; c_j $
with eigenvalues $\psi_j$ and $\bar\psi_j$
\begin{eqnarray}
\label{CPR}
c_{2j}\mid \psi_{2j}\rangle =- \psi_{2j}\mid \psi_{2j}\rangle &,&
\langle\bar{\psi}_{2j} \mid c^+_{2j} =
-\langle\bar{\psi}_{2j} \mid \bar{\psi}_{2j},\\
c^+_{2j+1}\mid \bar{\psi}_{2j+1}\rangle =
\bar{\psi}_{2j+1}\mid \bar{\psi}_{2j+1}\rangle &,&
\langle\psi_{2j+1}\mid c_{2j+1} =- \langle\psi_{2j+1}\mid \psi_{2j+1}.\nn
\end{eqnarray}

It is easy to calculate the scalar product of this states
\begin{eqnarray}
\label{TpR}
\langle \bar{\psi}_{2j}\mid \psi_{2j}\rangle=
e^{\bar{\psi}_{2j} \psi_{2j}},\qquad
\langle\psi_{2j+1} \mid \bar{\psi}_{2j+1}\rangle=
e^{\bar{\psi}_{2j+1} \psi_{2j+1}},
\end{eqnarray}
and find the completeness relations
\begin{eqnarray}
\int d\bar{\psi}_{2j}d\psi_{2j}\mid \psi_{2j}
\rangle \langle\bar{\psi}_{2j}\mid e^{\psi_{2j}\bar{\psi}_{2j}}&=& 1,
\nn\\
\label{CRE}
\int d\bar{\psi}_{2j+1}d\psi_{2j+1}\mid \bar{\psi}_{2j+1}\rangle
\langle\psi_{2j+1}\mid e^{\psi_{2j+1}\bar{\psi}_{2j+1}}&=& 1.
\end{eqnarray}

Let us  now pass to the coherent basis (\ref{CS1}, \ref{CS2}) in
the space of states $\prod_j V_j$ of the chain and calculate the
matrix elements of the $R^{\pm}_{2j,2j\pm1}$-operators between the
initial $\mid \psi_{2j}\rangle \in V_{2j} ,\;\;\;
\\
\mid \bar{\psi}_{2j\pm1}\rangle \in V_{2j\pm1} $ and final
$\langle\bar\psi_{2j}^{\pr}\mid \in V^{\pr}_{2j},
\;\;\langle\psi_{2j\pm1}^{\pr}\mid \in V_{2j\pm1}^{\pr}$ states.
By use of properties of coherent states it is easy to find from
the formula (\ref{dXX}), that
\begin{eqnarray}
\label{Rpsi}
&&R_{\psi_{2j},\bar{\psi}_{2j\pm1}}^{\bar{\psi}^{\pr}_{2j},\psi^{\pr}_{2j\pm1}}=
\langle\psi^{\pr}_{2j\pm1},\bar{\psi}^{\pr}_{2j}\mid\bar{R}^{\pm}_{2j,2j\pm1}
\mid\psi_{2j},\bar{\psi}_{2j\pm1}\rangle=\nn\\
&=&e^{\left[e^{i\alpha}a_{\pm}\bar{\psi}^{\pr}_{2j}\psi_{2j}
+e^{i\alpha^{\prime}}a_{\pm}\bar{\psi}_{2j\pm1}\psi^{\pr}_{2j\pm1}
-e^{i\alpha} b_{\pm}\bar{\psi}_{2j\pm1}\psi_{2j} +
e^{i\alpha^{\prime}} b_{\pm} \bar{\psi}^{\prime}_{2j}\psi^{\prime}_{2j\pm1}\right]}
\nn\\
&=& e^{-{\cal S}^{\pm}(\psi_{2j},\bar{\psi}_{2j\pm1},
\bar{\psi}^{\pr}_{2j},\psi^{\pr}_{2j\pm1})}.
\end{eqnarray}
This formula clarifies the convenience of introduction of the
solid lines in the picture for $R$-matrix (Fig.1) since the
parameters $a$ and $b$ getting a meaning of hopping parameters for
fermions on it.

The completeness relations (\ref{CRE}) define the multiplication
rule of $R$-operators (\ref{Rpsi}) and one can now express the
Partition Function $Z$ of the model before the disorder
calculations as  functional integral over the classical Grassmann
fields $\{\psi\}$
\begin{equation}
\label{Z}
\ZZ=\cT r T^N = \int\cD\{\bar\psi\} \cD\{\psi\}
e^{-\sum_{R-matrices}{\cal S}^{\pm}(\psi_{2j},\bar{\psi}_{2j\pm1},
\bar{\psi}^{\prime}_{2j},\psi^{\prime}_{2j\pm1}) + \sum_j \psi_j\bar{\psi}_j}.
\end{equation}
with the action defined on the $ML$ (see \cite{S,S1} for details).
We would like to emphasize now , that though we have started with
the $R$-matrix of the $XX$ model (or, which is the same, $2D$
Ising model), but the theory we have formulated is essentially
different from $XX$ model, since the $R$-matrices are disposed in
the Transfer Matrix (or in the  action) inhomogeneously.

In order to analyze the localization properties of the $CC$ model
we will consider now the dimensionless Landauer resistance, which
is the ratio of reflection over the transmission coefficients and
is nothing but the particular matrix elements of the direct
product of the Transfer Matrix $T$ with its Hermitian conjugate
$T^{\dagger}$. It was argued in the articles \cite{ATAF} that the
average of Landauer resistance defines the double of inverse of
the scaling localization length. Since each phase factor appears
locally only in one $R$-matrix and the disorder is full(there are
no correlations between different points
 in the distribution of the phase factors),  it is
clear from the formulas (\ref{MRR}) that
\begin{eqnarray}
\label{ZZR} \langle T \otimes T^{\dagger}\rangle=\left( \prod_j
\langle \bar{R^+}_{2j,2j+1}\otimes
(\bar{R^+}_{2j,2j+1})^{\dagger}\rangle \prod_i \langle
\check{R}^-_{2i-1,2i}\otimes
(\check{R}^-_{2i-1,2i})^{\dagger}\rangle\right)^N
\end{eqnarray}
This average is
easy to calculate in the $\psi$-basis of coherent states and we
need to introduce two copies of Grassmann fields , say $\psi_{\uparrow}$ and
 $\psi_{\downarrow}$, for $T$ and $T^{\dagger}$ respectively. The effective theory
is convenient to represented graphically on double §ML§, as in Figure 3,
expressing the hoppings of fermions separately.

We think that the Hermitian conjugation in the formula (\ref{ZZR})
can be defined in a generalized way. Namely, in $R^{\dagger}$ the
phases $e^{i\alpha}$ of the links in the expression (\ref{Rpsi})
of $R$-matrix can be changed by $e^{\frac{\phi}{4}-i\alpha}$ (
rather than by $e^{-i\alpha}$),  expressing the possibility that
our system is in vacuum  $\phi$-flux background.
\begin{figure}[ht]
\begin{center}
\setlength{\unitlength}{0.00087489in}
\begingroup\makeatletter\ifx\SetFigFont\undefined%
\gdef\SetFigFont#1#2#3#4#5{%
  \reset@font\fontsize{#1}{#2pt}%
  \fontfamily{#3}\fontseries{#4}\fontshape{#5}%
  \selectfont}%
\fi\endgroup%
{\renewcommand{\dashlinestretch}{30}
\begin{picture}(6762,5121)(0,-10)
\drawline(405,4644)(405,4644)
\drawline(855,4644)(855,4644)
\path(855,5094)(855,594)
\path(1755,5094)(1755,594)
\path(2655,5094)(2655,594)
\path(4455,5094)(4455,594)
\path(5355,5094)(5355,594)
\path(6255,5094)(6255,594)
\drawline(1305,4644)(1305,4644)
\drawline(1305,4644)(1305,4644)
\path(1755,4644)(2205,4644)
\path(2085.000,4614.000)(2205.000,4644.000)(2085.000,4674.000)
\path(2655,4644)(3105,4644)
\path(2985.000,4614.000)(3105.000,4644.000)(2985.000,4674.000)
\path(4455,4644)(4905,4644)
\path(4785.000,4614.000)(4905.000,4644.000)(4785.000,4674.000)
\path(5355,4644)(5805,4644)
\path(5685.000,4614.000)(5805.000,4644.000)(5685.000,4674.000)
\path(855,2844)(1305,2844)
\path(1185.000,2814.000)(1305.000,2844.000)(1185.000,2874.000)
\path(1755,2844)(2205,2844)
\path(2085.000,2814.000)(2205.000,2844.000)(2085.000,2874.000)
\path(2655,2844)(3105,2844)
\path(2985.000,2814.000)(3105.000,2844.000)(2985.000,2874.000)
\path(4455,2844)(4905,2844)
\path(4785.000,2814.000)(4905.000,2844.000)(4785.000,2874.000)
\path(5355,2844)(5805,2844)
\path(5685.000,2814.000)(5805.000,2844.000)(5685.000,2874.000)
\path(855,1044)(1305,1044)
\path(1185.000,1014.000)(1305.000,1044.000)(1185.000,1074.000)
\path(1755,1044)(2205,1044)
\path(2085.000,1014.000)(2205.000,1044.000)(2085.000,1074.000)
\path(2655,1044)(3105,1044)
\path(2985.000,1014.000)(3105.000,1044.000)(2985.000,1074.000)
\path(4455,1044)(4905,1044)
\path(4785.000,1014.000)(4905.000,1044.000)(4785.000,1074.000)
\drawline(5580,1044)(5580,1044)
\path(855,3744)(855,4194)
\path(885.000,4074.000)(855.000,4194.000)(825.000,4074.000)
\path(855,2844)(855,3294)
\path(885.000,3174.000)(855.000,3294.000)(825.000,3174.000)
\path(855,1944)(855,2394)
\path(885.000,2274.000)(855.000,2394.000)(825.000,2274.000)
\path(855,1044)(855,1494)
\path(885.000,1374.000)(855.000,1494.000)(825.000,1374.000)
\path(855,4644)(855,5094)
\path(885.000,4974.000)(855.000,5094.000)(825.000,4974.000)
\path(1755,4644)(1755,4194)
\path(1725.000,4314.000)(1755.000,4194.000)(1785.000,4314.000)
\path(1755,3744)(1755,3294)
\path(1725.000,3414.000)(1755.000,3294.000)(1785.000,3414.000)
\path(1755,2844)(1755,2394)
\path(1725.000,2514.000)(1755.000,2394.000)(1785.000,2514.000)
\path(1755,1944)(1755,1494)
\path(1725.000,1614.000)(1755.000,1494.000)(1785.000,1614.000)
\path(1755,1044)(1755,594)
\path(1725.000,714.000)(1755.000,594.000)(1785.000,714.000)
\path(2655,3744)(2655,4194)
\path(2685.000,4074.000)(2655.000,4194.000)(2625.000,4074.000)
\path(2655,4644)(2655,5094)
\path(2685.000,4974.000)(2655.000,5094.000)(2625.000,4974.000)
\path(2655,2844)(2655,3294)
\path(2685.000,3174.000)(2655.000,3294.000)(2625.000,3174.000)
\path(2655,1944)(2655,2394)
\path(2685.000,2274.000)(2655.000,2394.000)(2625.000,2274.000)
\path(2655,1044)(2655,1494)
\path(2685.000,1374.000)(2655.000,1494.000)(2625.000,1374.000)
\path(4455,1044)(4455,1494)
\path(4485.000,1374.000)(4455.000,1494.000)(4425.000,1374.000)
\path(4455,1944)(4455,2394)
\path(4485.000,2274.000)(4455.000,2394.000)(4425.000,2274.000)
\path(4455,2844)(4455,3294)
\path(4485.000,3174.000)(4455.000,3294.000)(4425.000,3174.000)
\path(4455,3744)(4455,4194)
\path(4485.000,4074.000)(4455.000,4194.000)(4425.000,4074.000)
\path(4455,4644)(4455,5094)
\path(4485.000,4974.000)(4455.000,5094.000)(4425.000,4974.000)
\path(5355,4644)(5355,4194)
\path(5325.000,4314.000)(5355.000,4194.000)(5385.000,4314.000)
\path(5355,3744)(5355,3294)
\path(5325.000,3414.000)(5355.000,3294.000)(5385.000,3414.000)
\path(5355,2844)(5355,2394)
\path(5325.000,2514.000)(5355.000,2394.000)(5385.000,2514.000)
\path(5355,1944)(5355,1494)
\path(5325.000,1614.000)(5355.000,1494.000)(5385.000,1614.000)
\path(5355,1044)(5355,594)
\path(5325.000,714.000)(5355.000,594.000)(5385.000,714.000)
\path(6255,3744)(6255,4194)
\path(6285.000,4074.000)(6255.000,4194.000)(6225.000,4074.000)
\path(6255,4644)(6255,5094)
\path(6285.000,4974.000)(6255.000,5094.000)(6225.000,4974.000)
\path(6255,2844)(6255,3294)
\path(6285.000,3174.000)(6255.000,3294.000)(6225.000,3174.000)
\path(6255,1944)(6255,2394)
\path(6285.000,2274.000)(6255.000,2394.000)(6225.000,2274.000)
\path(6255,1044)(6255,1494)
\path(6285.000,1374.000)(6255.000,1494.000)(6225.000,1374.000)
\path(6255,594)(6255,1044)
\path(6285.000,924.000)(6255.000,1044.000)(6225.000,924.000)
\path(1755,3744)(1305,3744)
\path(1425.000,3774.000)(1305.000,3744.000)(1425.000,3714.000)
\path(855,3744)(405,3744)
\path(525.000,3774.000)(405.000,3744.000)(525.000,3714.000)
\path(2655,3744)(2205,3744)
\path(2325.000,3774.000)(2205.000,3744.000)(2325.000,3714.000)
\path(4455,3744)(4005,3744)
\path(4125.000,3774.000)(4005.000,3744.000)(4125.000,3714.000)
\path(5355,3744)(4905,3744)
\path(5025.000,3774.000)(4905.000,3744.000)(5025.000,3714.000)
\path(6255,3744)(5805,3744)
\path(5925.000,3774.000)(5805.000,3744.000)(5925.000,3714.000)
\path(855,1944)(405,1944)
\path(525.000,1974.000)(405.000,1944.000)(525.000,1914.000)
\path(1755,1944)(1305,1944)
\path(1425.000,1974.000)(1305.000,1944.000)(1425.000,1914.000)
\path(2655,1944)(2205,1944)
\path(2325.000,1974.000)(2205.000,1944.000)(2325.000,1914.000)
\path(4455,1944)(4005,1944)
\path(4125.000,1974.000)(4005.000,1944.000)(4125.000,1914.000)
\path(5355,1944)(4905,1944)
\path(5025.000,1974.000)(4905.000,1944.000)(5025.000,1914.000)
\path(6255,1944)(5805,1944)
\path(5925.000,1974.000)(5805.000,1944.000)(5925.000,1914.000)
\dashline{60.000}(855,2844)(1305,3294)
\path(1241.360,3187.934)(1305.000,3294.000)(1198.934,3230.360)
\dashline{60.000}(1755,3744)(1305,3294)
\path(1368.640,3400.066)(1305.000,3294.000)(1411.066,3357.640)
\dashline{60.000}(1305,3294)(855,3744)
\path(961.066,3680.360)(855.000,3744.000)(918.640,3637.934)
\dashline{60.000}(1305,3294)(1755,2844)
\path(1648.934,2907.640)(1755.000,2844.000)(1691.360,2950.066)
\dashline{60.000}(1755,2844)(2205,2394)
\path(2098.934,2457.640)(2205.000,2394.000)(2141.360,2500.066)
\dashline{60.000}(2655,1944)(2205,2394)
\path(2311.066,2330.360)(2205.000,2394.000)(2268.640,2287.934)
\dashline{60.000}(1305,1494)(855,1944)
\path(961.066,1880.360)(855.000,1944.000)(918.640,1837.934)
\dashline{60.000}(1305,1494)(1755,1044)
\path(1648.934,1107.640)(1755.000,1044.000)(1691.360,1150.066)
\dashline{60.000}(855,1044)(1305,1494)
\path(1241.360,1387.934)(1305.000,1494.000)(1198.934,1430.360)
\dashline{60.000}(1755,1944)(1305,1494)
\path(1368.640,1600.066)(1305.000,1494.000)(1411.066,1557.640)
\dashline{60.000}(2205,2394)(1755,1944)
\path(1818.640,2050.066)(1755.000,1944.000)(1861.066,2007.640)
\dashline{60.000}(2205,2394)(2655,2844)
\path(2591.360,2737.934)(2655.000,2844.000)(2548.934,2780.360)
\dashline{60.000}(1755,4644)(2205,4194)
\path(2098.934,4257.640)(2205.000,4194.000)(2141.360,4300.066)
\dashline{60.000}(2205,4194)(1755,3744)
\path(1818.640,3850.066)(1755.000,3744.000)(1861.066,3807.640)
\dashline{60.000}(2205,4194)(2655,4644)
\path(2591.360,4537.934)(2655.000,4644.000)(2548.934,4580.360)
\dashline{60.000}(2655,3744)(2205,4194)
\path(2311.066,4130.360)(2205.000,4194.000)(2268.640,4087.934)
\dashline{60.000}(4905,1494)(4455,1944)
\path(4561.066,1880.360)(4455.000,1944.000)(4518.640,1837.934)
\dashline{60.000}(4905,1494)(5355,1044)
\path(5248.934,1107.640)(5355.000,1044.000)(5291.360,1150.066)
\dashline{60.000}(4905,3294)(4455,3744)
\path(4561.066,3680.360)(4455.000,3744.000)(4518.640,3637.934)
\dashline{60.000}(4905,3294)(5355,2844)
\path(5248.934,2907.640)(5355.000,2844.000)(5291.360,2950.066)
\dashline{60.000}(5355,2844)(5805,2394)
\path(5698.934,2457.640)(5805.000,2394.000)(5741.360,2500.066)
\dashline{60.000}(6255,1944)(5805,2394)
\path(5911.066,2330.360)(5805.000,2394.000)(5868.640,2287.934)
\dashline{60.000}(4455,2844)(4905,3294)
\path(4841.360,3187.934)(4905.000,3294.000)(4798.934,3230.360)
\dashline{60.000}(5355,3744)(4905,3294)
\path(4968.640,3400.066)(4905.000,3294.000)(5011.066,3357.640)
\dashline{60.000}(5805,4194)(5355,3744)
\path(5418.640,3850.066)(5355.000,3744.000)(5461.066,3807.640)
\dashline{60.000}(5805,4194)(6255,4644)
\put(6405,1654){\makebox(0,0)[lb]{\smash{{{\SetFigFont{14}{16.8}{\rmdefault}{\mddefault}
{\updefault}$T_2$}}}}}
\put(4205,3894){\makebox(0,0)[lb]{\smash{{{\SetFigFont{14}{16.8}{\rmdefault}{\mddefault}
{\updefault}$T_2$}}}}}
\put(6405,3454){\makebox(0,0)[lb]{\smash{{{\SetFigFont{14}{16.8}{\rmdefault}{\mddefault}
{\updefault}$T_1$}}}}}
\put(5105,4794){\makebox(0,0)[lb]{\smash{{{\SetFigFont{14}{16.8}{\rmdefault}{\mddefault}
{\updefault}$T_1$}}}}}
\path(6191.360,4537.934)(6255.000,4644.000)(6148.934,4580.360)
\dashline{60.000}(5355,4644)(5805,4194)
\path(5698.934,4257.640)(5805.000,4194.000)(5741.360,4300.066)
\dashline{60.000}(6255,3744)(5805,4194)
\path(5911.066,4130.360)(5805.000,4194.000)(5868.640,4087.934)
\dashline{60.000}(5805,2394)(5355,1944)
\path(5418.640,2050.066)(5355.000,1944.000)(5461.066,2007.640)
\dashline{60.000}(5355,1944)(4905,1494)
\path(4968.640,1600.066)(4905.000,1494.000)(5011.066,1557.640)
\dashline{60.000}(4455,1044)(4905,1494)
\path(4841.360,1387.934)(4905.000,1494.000)(4798.934,1430.360)
\dashline{60.000}(5805,2394)(6255,2844)
\path(6191.360,2737.934)(6255.000,2844.000)(6148.934,2780.360)
\dashline{60.000}(5355,1044)(5805,1044)
\path(5685.000,1014.000)(5805.000,1044.000)(5685.000,1074.000)
\path(945,5094)(945,594) \path(2745,5094)(2745,594)
\path(4545,5094)(4545,594) \path(5445,5094)(5445,594)
\dottedline{45}(1215,4554)(1395,4554)
\path(1275.000,4524.000)(1395.000,4554.000)(1275.000,4584.000)
\dottedline{45}(1215,4644)(1395,4644)
\path(1275.000,4614.000)(1395.000,4644.000)(1275.000,4674.000)
\dottedline{45}(2115,4554)(2205,4554)
\path(2085.000,4524.000)(2205.000,4554.000)(2085.000,4584.000)
\dottedline{45}(3015,4554)(3105,4554)
\path(2985.000,4524.000)(3105.000,4554.000)(2985.000,4584.000)
\drawline(4815,4554)(4815,4554)
\dottedline{45}(4815,4554)(4905,4554)
\path(4785.000,4524.000)(4905.000,4554.000)(4785.000,4584.000)
\dottedline{45}(5715,4554)(5805,4554)
\path(5685.000,4524.000)(5805.000,4554.000)(5685.000,4584.000)
\drawline(945,4104)(945,4104) \drawline(945,4104)(945,4104)
\dottedline{45}(945,4014)(945,4194)
\path(975.000,4074.000)(945.000,4194.000)(915.000,4074.000)
\dottedline{45}(1845,4284)(1845,4194)
\path(1815.000,4314.000)(1845.000,4194.000)(1875.000,4314.000)
\dottedline{45}(2745,4014)(2745,4104)
\path(2775.000,3984.000)(2745.000,4104.000)(2715.000,3984.000)
\dottedline{45}(4545,4014)(4545,4194)
\path(4575.000,4074.000)(4545.000,4194.000)(4515.000,4074.000)
\dottedline{45}(5445,4284)(5445,4194)
\path(5415.000,4314.000)(5445.000,4194.000)(5475.000,4314.000)
\drawline(6345,4104)(6345,4104)
\dottedline{45}(6345,4104)(6345,4194)
\path(6375.000,4074.000)(6345.000,4194.000)(6315.000,4074.000)
\drawline(1395,3654)(1395,3654) \drawline(1395,3654)(1395,3654)
\dottedline{45}(1485,3654)(1395,3654)
\path(1515.000,3684.000)(1395.000,3654.000)(1515.000,3624.000)
\dottedline{45}(2385,3654)(2295,3654)
\path(2415.000,3684.000)(2295.000,3654.000)(2415.000,3624.000)
\dottedline{45}(4095,3654)(4005,3654)
\path(4125.000,3684.000)(4005.000,3654.000)(4125.000,3624.000)
\drawline(4995,3654)(4995,3654)
\dottedline{45}(4995,3654)(4905,3654)
\path(5025.000,3684.000)(4905.000,3654.000)(5025.000,3624.000)
\dottedline{45}(5985,3654)(5895,3654)
\path(6015.000,3684.000)(5895.000,3654.000)(6015.000,3624.000)
\drawline(6345,3204)(6345,3204)
\dottedline{45}(6345,3204)(6345,3294)
\path(6375.000,3174.000)(6345.000,3294.000)(6315.000,3174.000)
\drawline(5445,3384)(5445,3384)
\dottedline{45}(5445,3384)(5445,3294)
\path(5415.000,3414.000)(5445.000,3294.000)(5475.000,3414.000)
\dottedline{45}(4545,3204)(4545,3294)
\path(4575.000,3174.000)(4545.000,3294.000)(4515.000,3174.000)
\drawline(2745,3204)(2745,3204)
\dottedline{45}(2745,3114)(2745,3204)
\path(2775.000,3084.000)(2745.000,3204.000)(2715.000,3084.000)
\drawline(1845,3204)(1845,3204)
\dottedline{45}(1845,3204)(1845,3114)
\path(1815.000,3234.000)(1845.000,3114.000)(1875.000,3234.000)
\dottedline{45}(945,3114)(945,3204)
\path(975.000,3084.000)(945.000,3204.000)(915.000,3084.000)
\dottedline{45}(1305,2754)(1395,2754)
\path(1275.000,2724.000)(1395.000,2754.000)(1275.000,2784.000)
\dottedline{45}(2115,2754)(2205,2754)
\path(2085.000,2724.000)(2205.000,2754.000)(2085.000,2784.000)
\dottedline{45}(3015,2754)(3105,2754)
\path(2985.000,2724.000)(3105.000,2754.000)(2985.000,2784.000)
\dottedline{45}(4815,2754)(4905,2754)
\path(4785.000,2724.000)(4905.000,2754.000)(4785.000,2784.000)
\dottedline{45}(5715,2754)(5895,2754)
\path(5775.000,2724.000)(5895.000,2754.000)(5775.000,2784.000)
\dottedline{45}(5985,1854)(5805,1854)
\path(5925.000,1884.000)(5805.000,1854.000)(5925.000,1824.000)
\dottedline{45}(4995,1854)(4905,1854)
\path(5025.000,1884.000)(4905.000,1854.000)(5025.000,1824.000)
\dottedline{45}(4095,1854)(4005,1854)
\path(4125.000,1884.000)(4005.000,1854.000)(4125.000,1824.000)
\dottedline{45}(2295,1854)(2205,1854)
\path(2325.000,1884.000)(2205.000,1854.000)(2325.000,1824.000)
\drawline(1395,1854)(1395,1854)
\dottedline{45}(1395,1854)(1305,1854)
\path(1425.000,1884.000)(1305.000,1854.000)(1425.000,1824.000)
\dottedline{45}(495,1854)(405,1854)
\path(525.000,1884.000)(405.000,1854.000)(525.000,1824.000)
\dottedline{45}(495,3654)(405,3654)
\path(525.000,3684.000)(405.000,3654.000)(525.000,3624.000)
\dottedline{45}(1215,954)(1305,954)
\path(1185.000,924.000)(1305.000,954.000)(1185.000,984.000)
\dottedline{45}(2115,954)(2205,954)
\path(2085.000,924.000)(2205.000,954.000)(2085.000,984.000)
\dottedline{45}(3015,954)(3105,954)
\path(2985.000,924.000)(3105.000,954.000)(2985.000,984.000)
\dottedline{45}(4815,954)(4905,954)
\path(4785.000,924.000)(4905.000,954.000)(4785.000,984.000)
\dottedline{45}(5715,954)(5805,954)
\path(5685.000,924.000)(5805.000,954.000)(5685.000,984.000)
\dottedline{45}(945,2304)(945,2394)
\path(975.000,2274.000)(945.000,2394.000)(915.000,2274.000)
\dottedline{45}(1845,2484)(1845,2304)
\path(1815.000,2424.000)(1845.000,2304.000)(1875.000,2424.000)
\dottedline{45}(2745,2304)(2745,2394)
\path(2775.000,2274.000)(2745.000,2394.000)(2715.000,2274.000)
\dottedline{45}(4545,2304)(4545,2394)
\path(4575.000,2274.000)(4545.000,2394.000)(4515.000,2274.000)
\dottedline{45}(5445,2484)(5445,2394)
\path(5415.000,2514.000)(5445.000,2394.000)(5475.000,2514.000)
\dottedline{45}(6345,2304)(6345,2394)
\path(6375.000,2274.000)(6345.000,2394.000)(6315.000,2274.000)
\dottedline{45}(6345,1404)(6345,1494)
\path(6375.000,1374.000)(6345.000,1494.000)(6315.000,1374.000)
\dottedline{45}(5445,1584)(5445,1494)
\path(5415.000,1614.000)(5445.000,1494.000)(5475.000,1614.000)
\dottedline{45}(4545,1404)(4545,1494)
\path(4575.000,1374.000)(4545.000,1494.000)(4515.000,1374.000)
\dottedline{45}(2745,1404)(2745,1494)
\path(2775.000,1374.000)(2745.000,1494.000)(2715.000,1374.000)
\dottedline{45}(1845,1674)(1845,1584)
\path(1815.000,1704.000)(1845.000,1584.000)(1875.000,1704.000)
\dottedline{45}(945,1404)(945,1494)
\path(975.000,1374.000)(945.000,1494.000)(915.000,1374.000)
\path(1845,5094)(1845,594) \path(6345,594)(6345,5094)(6345,5049)
\path(405,4644)(3060,4644) \path(450,4554)(3060,4554)
\path(450,3654)(3105,3654) \path(450,3744)(3105,3744)
\path(405,2844)(3060,2844) \path(405,2754)(3060,2754)
\path(450,1854)(3105,1854) \path(495,1944)(3105,1944)
\path(405,954)(3105,954) \path(405,1044)(3060,1044)
\path(4005,4644)(6705,4644) \path(4005,4554)(6705,4554)
\path(4095,3744)(6705,3744) \path(4095,3654)(6705,3654)
\path(4005,2844)(6705,2844) \path(4050,2754)(6705,2754)
\path(4005,1944)(6705,1944) \path(4005,1854)(6705,1854)
\path(4005,1044)(6705,1044) \path(4005,954)(6660,954)
\path(6570,954)(6705,954)
\path(6585.000,924.000)(6705.000,954.000)(6585.000,984.000)
\path(6570,1044)(6705,1044)
\path(6585.000,1014.000)(6705.000,1044.000)(6585.000,1074.000)
\path(6570,2754)(6705,2754)
\path(6585.000,2724.000)(6705.000,2754.000)(6585.000,2784.000)
\path(6525,2844)(6705,2844)
\path(6585.000,2814.000)(6705.000,2844.000)(6585.000,2874.000)
\path(6570,4554)(6660,4554)
\path(6540.000,4524.000)(6660.000,4554.000)(6540.000,4584.000)
\path(6525,4644)(6750,4644)
\path(6630.000,4614.000)(6750.000,4644.000)(6630.000,4674.000)
\put(0,1044){\makebox(0,0)[lb]{\smash{{{\SetFigFont{12}{14.4}{\rmdefault}{\mddefault}
{\updefault}2j-1}}}}}
\put(0,1989){\makebox(0,0)[lb]{\smash{{{\SetFigFont{12}{14.4}{\rmdefault}{\mddefault}
{\updefault}2j}}}}}
\put(0,2889){\makebox(0,0)[lb]{\smash{{{\SetFigFont{12}{14.4}{\rmdefault}{\mddefault}
{\updefault}2j+1}}}}}
\put(0,3789){\makebox(0,0)[lb]{\smash{{{\SetFigFont{12}{14.4}{\rmdefault}{\mddefault}
{\updefault}2j+2}}}}}
\put(2205,369){\makebox(0,0)[lb]{\smash{{{\SetFigFont{14}{16.8}{\rmdefault}{\mddefault}
{\updefault}$M_2$}}}}}
\put(4905,369){\makebox(0,0)[lb]{\smash{{{\SetFigFont{14}{16.8}{\rmdefault}{\mddefault}
{\updefault}$M_1$}}}}}
\put(5805,369){\makebox(0,0)[lb]{\smash{{{\SetFigFont{14}{16.8}{\rmdefault}{\mddefault}
{\updefault}$M_2$}}}}}
\put(1305,369){\makebox(0,0)[lb]{\smash{{{\SetFigFont{14}{16.8}{\rmdefault}{\mddefault}
{\updefault}$M_1$}}}}}
\put(2205,2934){\makebox(0,0)[lb]{\smash{{{\SetFigFont{12}{14.4}{\rmdefault}{\bfdefault}
{\updefault}$e^{\frac{i\pi}{4}}b$}}}}}
\put(2205,1674){\makebox(0,0)[lb]{\smash{{{\SetFigFont{12}{14.4}{\rmdefault}{\bfdefault}
{\updefault}$e^{\frac{i\pi}{4}}b$}}}}}
\put(1215,2034){\makebox(0,0)[lb]{\smash{{{\SetFigFont{12}{14.4}{\rmdefault}{\bfdefault}
{\updefault}$e^{\frac{i\pi}{4}}a$}}}}}
\put(1215,2474){\makebox(0,0)[lb]{\smash{{{\SetFigFont{12}{14.4}{\rmdefault}{\bfdefault}
{\updefault}$e^{\frac{i\pi}{4}}a$}}}}}
\put(1215,3834){\makebox(0,0)[lb]{\smash{{{\SetFigFont{12}{14.4}{\rmdefault}{\bfdefault}
{\updefault}$e^{\frac{i\pi}{4}}a$}}}}}
\put(2880,2304){\makebox(0,0)[lb]{\smash{{{\SetFigFont{12}{14.4}{\rmdefault}{\bfdefault}
{\updefault}$e^{\frac{i\pi}{4}}a$}}}}}
\put(2475,1404){\makebox(0,0)[lb]{\smash{{{\SetFigFont{12}{14.4}{\rmdefault}{\bfdefault}
{\updefault}$b$}}}}}
\put(2475,3204){\makebox(0,0)[lb]{\smash{{{\SetFigFont{12}{14.4}{\rmdefault}{\bfdefault}
{\updefault}$b$}}}}}
\put(1215,4734){\makebox(0,0)[lb]{\smash{{{\SetFigFont{12}{14.4}{\rmdefault}{\bfdefault}
{\updefault}$e^{\frac{i\pi}{4}}a$}}}}}
\put(4815,684){\makebox(0,0)[lb]{\smash{{{\SetFigFont{12}{14.4}{\rmdefault}{\bfdefault}
{\updefault}$a$}}}}}
\put(4815,2034){\makebox(0,0)[lb]{\smash{{{\SetFigFont{12}{14.4}{\rmdefault}{\bfdefault}
{\updefault}$a$}}}}}
\put(4185,2304){\makebox(0,0)[lb]{\smash{{{\SetFigFont{12}{14.4}{\rmdefault}{\bfdefault}
{\updefault}$a$}}}}}
\put(4815,2529){\makebox(0,0)[lb]{\smash{{{\SetFigFont{12}{14.4}{\rmdefault}{\bfdefault}
{\updefault}$a$}}}}}
\put(5125,2304){\makebox(0,0)[lb]{\smash{{{\SetFigFont{12}{14.4}{\rmdefault}{\bfdefault}
{\updefault}$-a$}}}}}
\put(4815,3834){\makebox(0,0)[lb]{\smash{{{\SetFigFont{12}{14.4}{\rmdefault}{\bfdefault}
{\updefault}$a$}}}}}
\put(4815,4779){\makebox(0,0)[lb]{\smash{{{\SetFigFont{12}{14.4}{\rmdefault}{\bfdefault}
{\updefault}$a$}}}}}
\put(5805,4734){\makebox(0,0)[lb]{\smash{{{\SetFigFont{12}{14.4}{\rmdefault}{\bfdefault}
{\updefault}$b$}}}}}
\put(5805,3429){\makebox(0,0)[lb]{\smash{{{\SetFigFont{12}{14.4}{\rmdefault}{\bfdefault}
{\updefault}$b$}}}}}
\put(5805,2934){\makebox(0,0)[lb]{\smash{{{\SetFigFont{12}{14.4}{\rmdefault}{\bfdefault}
{\updefault}$b$}}}}}
\put(5805,1674){\makebox(0,0)[lb]{\smash{{{\SetFigFont{12}{14.4}{\rmdefault}{\bfdefault}
{\updefault}$b$}}}}}
\put(5805,684){\makebox(0,0)[lb]{\smash{{{\SetFigFont{12}{14.4}{\rmdefault}{\bfdefault}
{\updefault}$b$}}}}}
\put(6425,2304){\makebox(0,0)[lb]{\smash{{{\SetFigFont{12}{14.4}{\rmdefault}{\bfdefault}
{\updefault}$-a$}}}}}
\put(6075,3204){\makebox(0,0)[lb]{\smash{{{\SetFigFont{12}{14.4}{\rmdefault}{\bfdefault}
{\updefault}$b$}}}}}
\put(5130,4104){\makebox(0,0)[lb]{\smash{{{\SetFigFont{12}{14.4}{\rmdefault}{\bfdefault}
{\updefault}$a$}}}}}
\put(6425,4104){\makebox(0,0)[lb]{\smash{{{\SetFigFont{12}{14.4}{\rmdefault}{\bfdefault}
{\updefault}$a$}}}}}
\put(4455,369){\makebox(0,0)[lb]{\smash{{{\SetFigFont{12}{14.4}{\rmdefault}{\bfdefault}
{\updefault}t}}}}}
\put(5355,369){\makebox(0,0)[lb]{\smash{{{\SetFigFont{12}{14.4}{\rmdefault}{\bfdefault}
{\updefault}t+1}}}}}
\put(6255,369){\makebox(0,0)[lb]{\smash{{{\SetFigFont{12}{14.4}{\rmdefault}{\bfdefault}
{\updefault}t+2}}}}}
\put(855,369){\makebox(0,0)[lb]{\smash{{{\SetFigFont{12}{14.4}{\rmdefault}{\bfdefault}
{\updefault}t}}}}}
\put(2655,369){\makebox(0,0)[lb]{\smash{{{\SetFigFont{12}{14.4}{\rmdefault}{\bfdefault}
{\updefault}t+2}}}}}
\put(5355,54){\makebox(0,0)[lb]{\smash{{{\SetFigFont{14}{16.8}{\rmdefault}{\bfdefault}
{\updefault}b)}}}}}
\put(1755,54){\makebox(0,0)[lb]{\smash{{{\SetFigFont{14}{16.8}{\rmdefault}{\bfdefault}
{\updefault}a)}}}}}
\put(1845,3204){\makebox(0,0)[lb]{\smash{{{\SetFigFont{12}{14.4}{\rmdefault}{\bfdefault}
{\updefault}$e^{\frac{i\pi}{4}}b$}}}}}
\put(2205,684){\makebox(0,0)[lb]{\smash{{{\SetFigFont{12}{14.4}{\rmdefault}{\bfdefault}
{\updefault}$e^{\frac{i\pi}{4}}b$}}}}}
\put(2205,3424){\makebox(0,0)[lb]{\smash{{{\SetFigFont{12}{14.4}{\rmdefault}{\bfdefault}
{\updefault}$e^{\frac{i\pi}{4}}b$}}}}}
\put(2205,4734){\makebox(0,0)[lb]{\smash{{{\SetFigFont{12}{14.4}{\rmdefault}{\bfdefault}
{\updefault}$e^{\frac{i\pi}{4}}b$}}}}}
\put(1035,1404){\makebox(0,0)[lb]{\smash{{{\SetFigFont{12}{14.4}{\rmdefault}{\bfdefault}
{\updefault}$b$}}}}}
\put(1835,1404){\makebox(0,0)[lb]{\smash{{{\SetFigFont{12}{14.4}{\rmdefault}{\bfdefault}
{\updefault}$e^{\frac{i\pi}{4}}b$}}}}}
\put(1530,1404){\makebox(0,0)[lb]{\smash{{{\SetFigFont{12}{14.4}{\rmdefault}{\bfdefault}
{\updefault}$b$}}}}}
\put(1215,1674){\makebox(0,0)[lb]{\smash{{{\SetFigFont{12}{14.4}{\rmdefault}{\bfdefault}
{\updefault}$a$}}}}}
\put(1215,2934){\makebox(0,0)[lb]{\smash{{{\SetFigFont{12}{14.4}{\rmdefault}{\bfdefault}
{\updefault}$a$}}}}}
\put(1215,3474){\makebox(0,0)[lb]{\smash{{{\SetFigFont{12}{14.4}{\rmdefault}{\bfdefault}
{\updefault}$a$}}}}}
\put(1215,4374){\makebox(0,0)[lb]{\smash{{{\SetFigFont{12}{14.4}{\rmdefault}{\bfdefault}
{\updefault}$a$}}}}}
\put(2205,4374){\makebox(0,0)[lb]{\smash{{{\SetFigFont{12}{14.4}{\rmdefault}{\bfdefault}
{\updefault}$b$}}}}}
\put(2205,3834){\makebox(0,0)[lb]{\smash{{{\SetFigFont{12}{14.4}{\rmdefault}{\bfdefault}
{\updefault}$b$}}}}}
\put(2205,2574){\makebox(0,0)[lb]{\smash{{{\SetFigFont{12}{14.4}{\rmdefault}{\bfdefault}
{\updefault}$b$}}}}}
\put(2205,2034){\makebox(0,0)[lb]{\smash{{{\SetFigFont{12}{14.4}{\rmdefault}{\bfdefault}
{\updefault}$b$}}}}}
\put(2205,1134){\makebox(0,0)[lb]{\smash{{{\SetFigFont{12}{14.4}{\rmdefault}{\bfdefault}
{\updefault}$b$}}}}}
\put(1215,1134){\makebox(0,0)[lb]{\smash{{{\SetFigFont{12}{14.4}{\rmdefault}{\bfdefault}
{\updefault}$a$}}}}}
\put(1215,684){\makebox(0,0)[lb]{\smash{{{\SetFigFont{12}{14.4}{\rmdefault}{\bfdefault}
{\updefault}$e^{\frac{i\pi}{4}}a$}}}}}
\put(1755,369){\makebox(0,0)[lb]{\smash{{{\SetFigFont{12}{14.4}{\rmdefault}{\bfdefault}
{\updefault}t+1}}}}}
\put(1035,2304){\makebox(0,0)[lb]{\smash{{{\SetFigFont{12}{14.4}{\rmdefault}{\bfdefault}
{\updefault}$a$}}}}}
\put(1035,3204){\makebox(0,0)[lb]{\smash{{{\SetFigFont{12}{14.4}{\rmdefault}{\bfdefault}
{\updefault}$b$}}}}}
\put(500,1404){\makebox(0,0)[lb]{\smash{{{\SetFigFont{12}{14.4}{\rmdefault}{\bfdefault}
{\updefault}$e^{\frac{i\pi}{4}}b$}}}}}
\put(500,2304){\makebox(0,0)[lb]{\smash{{{\SetFigFont{12}{14.4}{\rmdefault}{\bfdefault}
{\updefault}$e^{\frac{i\pi}{4}}a$}}}}}
\put(500,3204){\makebox(0,0)[lb]{\smash{{{\SetFigFont{12}{14.4}{\rmdefault}{\bfdefault}
{\updefault}$e^{\frac{i\pi}{4}}b$}}}}}
\put(500,4104){\makebox(0,0)[lb]{\smash{{{\SetFigFont{12}{14.4}{\rmdefault}{\bfdefault}
{\updefault}$e^{\frac{i\pi}{4}}a$}}}}}
\put(1035,4104){\makebox(0,0)[lb]{\smash{{{\SetFigFont{12}{14.4}{\rmdefault}{\bfdefault}
{\updefault}$a$}}}}}
\put(1935,4104){\makebox(0,0)[lb]{\smash{{{\SetFigFont{12}{14.4}{\rmdefault}{\bfdefault}
{\updefault}$a$}}}}}
\put(1935,2304){\makebox(0,0)[lb]{\smash{{{\SetFigFont{12}{14.4}{\rmdefault}{\bfdefault}
{\updefault}$a$}}}}}
\put(2880,1404){\makebox(0,0)[lb]{\smash{{{\SetFigFont{12}{14.4}{\rmdefault}{\bfdefault}
{\updefault}$e^{\frac{i\pi}{4}}b$}}}}}
\put(2880,4104){\makebox(0,0)[lb]{\smash{{{\SetFigFont{12}{14.4}{\rmdefault}{\bfdefault}
{\updefault}$e^{\frac{i\pi}{4}}a$}}}}}
\put(2475,4104){\makebox(0,0)[lb]{\smash{{{\SetFigFont{12}{14.4}{\rmdefault}{\bfdefault}
{\updefault}$a$}}}}}
\put(1330,2254){\makebox(0,0)[lb]{\smash{{{\SetFigFont{12}{14.4}{\rmdefault}{\bfdefault}
{\updefault}$e^{\frac{i\pi}{4}}a$}}}}}
\put(1530,3204){\makebox(0,0)[lb]{\smash{{{\SetFigFont{12}{14.4}{\rmdefault}{\bfdefault}
{\updefault}$b$}}}}}
\put(1330,4104){\makebox(0,0)[lb]{\smash{{{\SetFigFont{12}{14.4}{\rmdefault}{\bfdefault}
{\updefault}$e^{\frac{i\pi}{4}}a$}}}}}
\put(2880,3204){\makebox(0,0)[lb]{\smash{{{\SetFigFont{12}{14.4}{\rmdefault}{\bfdefault}
{\updefault}$e^{\frac{i\pi}{4}}b$}}}}}
\put(4185,3204){\makebox(0,0)[lb]{\smash{{{\SetFigFont{12}{14.4}{\rmdefault}{\bfdefault}
{\updefault}$-b$}}}}}
\put(4185,4104){\makebox(0,0)[lb]{\smash{{{\SetFigFont{12}{14.4}{\rmdefault}{\bfdefault}
{\updefault}$-a$}}}}}
\put(4635,3204){\makebox(0,0)[lb]{\smash{{{\SetFigFont{12}{14.4}{\rmdefault}{\bfdefault}
{\updefault}$b$}}}}}
\put(4635,2304){\makebox(0,0)[lb]{\smash{{{\SetFigFont{12}{14.4}{\rmdefault}{\bfdefault}
{\updefault}$a$}}}}}
\put(4635,1404){\makebox(0,0)[lb]{\smash{{{\SetFigFont{12}{14.4}{\rmdefault}{\bfdefault}
{\updefault}$b$}}}}}
\put(4635,4104){\makebox(0,0)[lb]{\smash{{{\SetFigFont{12}{14.4}{\rmdefault}{\bfdefault}
{\updefault}$a$}}}}}
\put(4815,3429){\makebox(0,0)[lb]{\smash{{{\SetFigFont{12}{14.4}{\rmdefault}{\bfdefault}
{\updefault}$a$}}}}}
\put(4815,4374){\makebox(0,0)[lb]{\smash{{{\SetFigFont{12}{14.4}{\rmdefault}{\bfdefault}
{\updefault}$a$}}}}}
\put(4815,1719){\makebox(0,0)[lb]{\smash{{{\SetFigFont{12}{14.4}{\rmdefault}{\bfdefault}
{\updefault}$a$}}}}}
\put(4815,2934){\makebox(0,0)[lb]{\smash{{{\SetFigFont{12}{14.4}{\rmdefault}{\bfdefault}
{\updefault}$a$}}}}}
\put(4770,1134){\makebox(0,0)[lb]{\smash{{{\SetFigFont{12}{14.4}{\rmdefault}{\bfdefault}
{\updefault}$a$}}}}}
\put(5130,3204){\makebox(0,0)[lb]{\smash{{{\SetFigFont{12}{14.4}{\rmdefault}{\bfdefault}
{\updefault}$-b$}}}}}
\put(5805,2574){\makebox(0,0)[lb]{\smash{{{\SetFigFont{12}{14.4}{\rmdefault}{\bfdefault}
{\updefault}$b$}}}}}
\put(5805,2034){\makebox(0,0)[lb]{\smash{{{\SetFigFont{12}{14.4}{\rmdefault}{\bfdefault}
{\updefault}$b$}}}}}
\put(5805,1134){\makebox(0,0)[lb]{\smash{{{\SetFigFont{12}{14.4}{\rmdefault}{\bfdefault}
{\updefault}$b$}}}}}
\put(5130,1404){\makebox(0,0)[lb]{\smash{{{\SetFigFont{12}{14.4}{\rmdefault}{\bfdefault}
{\updefault}$b$}}}}}
\put(5580,1404){\makebox(0,0)[lb]{\smash{{{\SetFigFont{12}{14.4}{\rmdefault}{\bfdefault}
{\updefault}$b$}}}}}
\put(5535,2304){\makebox(0,0)[lb]{\smash{{{\SetFigFont{12}{14.4}{\rmdefault}{\bfdefault}
{\updefault}$a$}}}}}
\put(5535,4104){\makebox(0,0)[lb]{\smash{{{\SetFigFont{12}{14.4}{\rmdefault}{\bfdefault}
{\updefault}$a$}}}}}
\put(6075,4104){\makebox(0,0)[lb]{\smash{{{\SetFigFont{12}{14.4}{\rmdefault}{\bfdefault}
{\updefault}$b$}}}}}
\put(5805,4374){\makebox(0,0)[lb]{\smash{{{\SetFigFont{12}{14.4}{\rmdefault}{\bfdefault}
{\updefault}$b$}}}}}
\put(5760,3789){\makebox(0,0)[lb]{\smash{{{\SetFigFont{12}{14.4}{\rmdefault}{\bfdefault}
{\updefault}$b$}}}}}
\put(4185,1404){\makebox(0,0)[lb]{\smash{{{\SetFigFont{12}{14.4}{\rmdefault}{\bfdefault}
{\updefault}$b$}}}}}
\put(6450,3204){\makebox(0,0)[lb]{\smash{{{\SetFigFont{12}{14.4}{\rmdefault}{\bfdefault}
{\updefault}$b$}}}}}
\put(5535,3204){\makebox(0,0)[lb]{\smash{{{\SetFigFont{12}{14.4}{\rmdefault}{\bfdefault}
{\updefault}$b$}}}}}
\put(2475,2304){\makebox(0,0)[lb]{\smash{{{\SetFigFont{12}{14.4}{\rmdefault}{\bfdefault}
{\updefault}$a$}}}}}
\put(6075,2304){\makebox(0,0)[lb]{\smash{{{\SetFigFont{12}{14.4}{\rmdefault}{\bfdefault}
{\updefault}$a$}}}}}
\put(6075,1404){\makebox(0,0)[lb]{\smash{{{\SetFigFont{12}{14.4}{\rmdefault}{\bfdefault}
{\updefault}$b$}}}}}
\put(6380,1404){\makebox(0,0)[lb]{\smash{{{\SetFigFont{12}{14.4}{\rmdefault}{\bfdefault}
{\updefault}$-b$}}}}}
\end{picture}}
\vspace{-0.7cm}
\end{center}
\caption{The $CC$ model after disorder was taken into account. The
hopping parameters of two type fermions are marked separately.
Pictures a) and b) differs by gauge transformation.}
\end{figure}
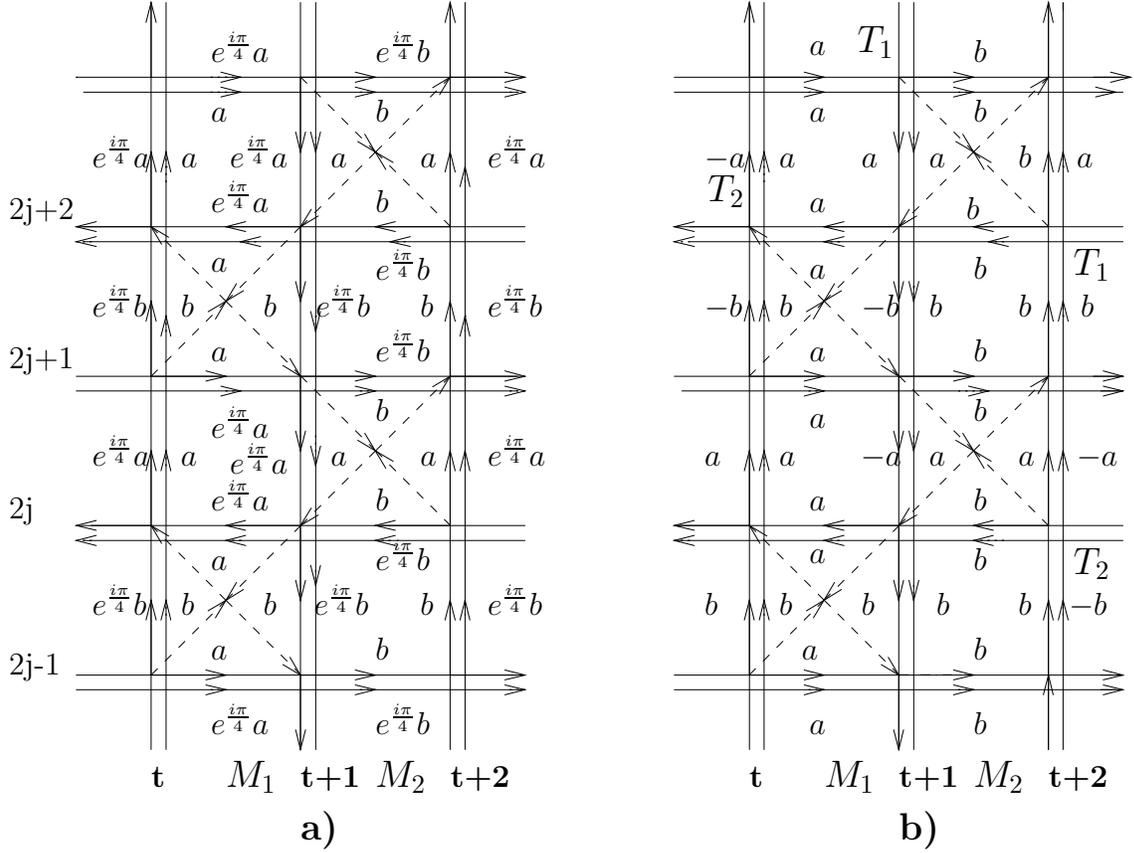
Then, the average
$\langle\check{R}_{2j,2j\pm1}\otimes\check{R}^{\dagger}_{2j,2j\pm1}
\rangle$ in case of Gaussian distribution $\cP(\{\alpha_j\})=
\prod_{j}\frac{1}{\kappa\sqrt\pi}
exp\left(-\frac{\alpha_{j}^2}{\kappa^2}\right)$ of phases can be
calculated easily and one will find the $R$-matrix of the Hubbard
model (see \cite{S1}). But we do not have ordinary Hubbard model
in a result, because , as it shown in Figure 3, in the Partition
Function two type of $R$-matrices  have to bee disposed in a
staggered way. We will write down now only the expression of
$R$-matrix for the background flux $\phi =\pi$ and in the strong
coupling limit ($\kappa \rightarrow \infty$), corresponding to
homogenous distribution of phases over the circle, because, as it
appeared, in this case the final model is integrable.
\begin{eqnarray}
\label{Hub}
(R^{\pm})_{\psi_{2j,\si},\bar{\psi}_{2j\pm1,\si}}^{\bar{\psi}^{\pr}_{2j,\si},
\psi^{\pr}_{2j\pm1,\si}}&=&
exp\left\{e^{i\frac{\pi}{4}}\bar{\psi}_{2j\pm1,\uparrow}
(a_{\pm}\psi^{\pr}_{2j\pm1,\uparrow}- b_{\pm}\psi_{2j,\uparrow})
(a_{\pm}\bar{\psi}_{2j\pm1,\downarrow}-
b_{\pm}\bar{\psi}^{\pr}_{2j,\downarrow})
\psi^{\pr}_{2j\pm1,\downarrow}\right.\nn\\
&+&\left.e^{i\frac{\pi}{4}}\bar{\psi}^{\pr}_{2j,\uparrow}
(a_{\pm}\psi_{2j,\uparrow}+ b_{\pm}\psi^{\pr}_{2j\pm1,\uparrow})
(a_{\pm}\bar{\psi}^{\pr}_{2j,\downarrow}+
b_{\pm}\bar{\psi}_{2j\pm1,\downarrow})\psi_{2j,\downarrow}
\right\}.
\end{eqnarray}
Due to rotational invariance one should take
$a_+=b_-=a,\;\;b_{+}=a_-=b$ and this two matrices have a chess
like disposition on the $ML$, as it is shown in the Figure 3a.

The $R$-matrices (\ref{Hub}) define an integrable model of the
type developed recently in a chain of articles \cite{APSS, S1}.
First, it is easy to make an $U(1)$ gauge transformation and pass
from the distribution of hopping parameters of the model as it
defined by the formula (\ref{Hub}) and shown on the Fig.3a, to the
one, which is drown on Fig.3b. This gauge transformation is
possible, because the distribution of fluxes through the
plaquettes of $ML$ is the same in both cases. Then, following
\cite{APSS}, let us define a new Transfer Matrix of the model as a
product of $R$-matrices along a chain, which is rotated on angle
$\pi/4$ with respect to vertical access. From the Fig.3b one can
see, that we have two different lines of Transfer Matrices
\begin{eqnarray}
\label{THub} T_1 =\prod_j R^-_{2j,2j+1}R^+_{2j-1,2j},\qquad T_2
=\prod_j R^{-,\iota}_{2j,2j+1}R^{+,\iota}_{2j-1,2j}.
\end{eqnarray}
The $R$-matrices of the first line (marked $T_1$ in the Fig.3b)
are defined by the formulas (\ref{Hub}) but without factors
$e^{i\frac{\pi}{4}}$ in all hopping parameters. In the second line
(marked $T_2$ in the Fig.3b) the $R$-matrices have minus sign in
front of vertical hopping parameters of one of fermions (say
spin-up), which we marked by the subscript $\iota$ in the
expression (\ref{THub}).
 But this is precisely the $\iota$ operation defined in the articles
\cite{APSS}. We have an inhomogeneous model. Besides of the
$\iota$ operation differing neighbor chains, the $R$ matrices are
staggered along of each chain. It was shown in \cite{APSS}, that
this $Z_2$ graded structure demands two sets of Yang-Baxter's
Equation ($YBE$) in order to ensure the commutativity of the
Transfer Matrices with different spectral parameters $u$ and $v$.
\begin{eqnarray}
\label{YBE} R_{23}(u,v)R^{+,\iota}_{12}(u)R^{-}_{23}(v)&=&
R^{-,\iota}_{1,2}(v)R^{+}_{2,3}(u)R^{\iota}_{1,2}(u,v),\nn\\
R^{\iota}_{2,3}(u,v)R^{-,\iota}_{1,2}(u)R^{+}_{2,3}(v)&=&
R^{+,\iota}_{1,2}(v)R^{-}_{2,3}(u)R_{1,2}(u,v).
\end{eqnarray}
One can check by direct calculations, that the $YBE$ (\ref{YBE})
for intertwinners $R_{i,i+1}(u,v)$ and $R^{-,\iota}_{i,i+1}$ have
a solution of the same form as (\ref{Hub}) with
\begin{eqnarray}
\label{sol} b^2(u,v) = (b^+(u))^2-(b^-(v))^2,\qquad a^2(u,v)=1-
b^2(u,v).
\end{eqnarray}
Therefore our model is integrable.  This nontrivial fact gives us
the possibility to use the powerful method of $ABA$  \cite{Bax,FT}
and investigate $QH$ plateau-plateau transitions by the exact
technique. This is the subject of further investigations.

In conclusion I would like to point out the main results of this
work. The exact Lagrangian formulation of the $CC$ model before
the disorder is taken into account is presented based on staggered
$XX$ model. After calculation of the average of the  Landauer
resistance with constant distribution of random phases over the
circle we have obtained a new type integrable model  with two
fermions (spin up/down). The integrability will allow to use the
powerful method of $ABA$ in order to investigate the
plateau-plateau transitions in $QHE$ exactly. If the bosonic
partner of this construction will be found one can also formulate
the corresponding supersymmetric model, which will allow to average
the free energy (rather than Landauer resistivity) of the $CC$
model.

 ACKNOWLEDGMENTS:
 The author wish to thank J.Ambjorn D.Arnaudon, A.Belavin, R.Flume,
T.Hakobyan,
D.Karakhanyan, R.Poghossian, V.Rittenberg, T.Sedrakyan, P.Sorba
and M.Zirnbauer for numerous and productive discussions. The work
was partially supported by the  SCOPE grant of SNF and INTAS grant
00-390.



\baselineskip=11pt
\begin{thebibliography}{10}

\bibitem{CC} J.Chalker, P.Coddington, J.Phys. {\bf C 21} (1988)
  2665.
\bibitem{LWK} Dung-Hai Lee, Ziqiang Wang, S. Kivelson,
Phys.Rev.Lett. {\bf 70}(1993) 4130.

\bibitem{ChoF1} S. Cho, M.P.A.Fisher, Phys.Rev.{\bf B 55}(1997)
1025.

\bibitem{BH} B. Huckenstein, Rev.Mod.Phys. {\bf 67} (1995) 357.

\bibitem{WEK} H.P. Wei et al., Phys.Rev.Lett.{\bf 61}(1988) 1297,\\
            L. Engel et al., Surf.Sci. {229}(1990)13,\\
            S. Koch et al., Phys.Rev.Lett.{\bf 67}(1991) 883.

\bibitem{McEuen} P.L. McEuen et al.,Phys.Rev.Lett.{\bf 64}(1990) 2062.

\bibitem{DHL1} Dung-Hai Lee, Phys.Rev.{\bf B 50}(1994)10788.

\bibitem{DHL2} Dung-Hai Lee, Ziqiang Wang, Phil.Mag.Lett.{\bf 73}(1996)145.

\bibitem{Kim} Y.-B. Kim, Phys.Rev.{\bf B 53}(1996)16420.

\bibitem{MT1} J. Kondev, J.B.Marston,  cond-mat/9612223,\\
               J.B.Marston, Shan-Wen Tsai, Phys.Rev.Lett.{\bf 82}(1999)4906.

\bibitem{Zirn1} M.R. Zirnbauer, Ann.d.Physik {\bf 3}(1994) 513.

\bibitem{KHAC} V.Kagalovsky, B.Horovitz, Y.Avishai, J.T.Chalker,
               Phys.Rev.Lett{\bf 82}(1999) 3516,\\
               T.Senthil, J.B.Marston, M.P.A.Fisher, Phys.Rev.{\bf B 60}(1999)4245.

\bibitem{GRS1} Ilya.A.Gruzberg, N.Read, Subir Sadchev,
Phys.Rev.{\bf B 55}(1997)10593,\\
Ilya.A.Gruzberg, Andreas W.W. Ludwig, N.Read,
Phys.Rev.Lett {\bf 82}(1999)4524.


\bibitem{AJS} Daniel P.Arovas, Martin Janssen, Boris Shapiro,
Phys.Rev.{\bf B 56}(1997)4751.

\bibitem{KM1} M. Klesse, M. Metzler, Europhys.Lett. {\bf 32}(1995) 229,\\
              M.Klesse, M.Metzler, Phys.Rev.Lett. {\bf 79}(1997) 721.

\bibitem{ChoF2} S.Cho, M.P.A.Fisher, Phys.Rev.{\bf B 55}(1997)1637.

\bibitem{RL}  N.Read,  A. W.W.Ludwig,  Phys.Rev. {\bf  B 63}(2001) 024404.

\bibitem{GRL} I.A.Gruzberg, N.Read,  A. W.W. Ludwig,
Phys.Rev. {\bf  B 63}(2001) 104422.

\bibitem{MCh1} F.Merz, J.Chalker, cond-mat/0106023.

\bibitem{BC} M.Bocquet, J.Chalker, cond-mat/0210695.

\bibitem{S} A.Sedrakyan, Nucl.Phys.{\bf 554 B [FS]} (1999) 514.

\bibitem{APSS}  D.Arnaudon, R.Poghossian, A.Sedrakyan, P.Sorba,
  Nucl.Phys.{\bf 588B}(2000) 638,\\
      T.Sedrakyan, Nucl. Phys. {\bf 608 B [FS]} (2001) 557-576,\\
       D.Arnaudon, A.Sedrakyan, T.Sedrakyan, P.Sorba,
        Lett. Math.Phys.{\bf 58}(2001)209.

\bibitem{S1} A.Sedrakyan, \textsl{
  Integrable Chain Models with Staggered $R$-matrices,}
  Contribution to the proceedings of Advanced NATO Workshop on Statistical
  Field Theories, Editors: A.Capelli, G.Mussardo, Como, June 18-23, 2001,
cond-mat/0112077.

\bibitem{ATAF}  P.W.Anderson, D.J.Thouless, E.Abrahams,
D.S.Fisher,  Phys.Rev. {\bf  B 22}(1980) 3519,\\
P.W.Anderson, Phys.Rev. {\bf  B 23}(1981) 4828.

\bibitem {Bax} J.Baxter, \textsl{Exactly Solved Models in Statistical
    Mechanics,} Academic Press,London (1989).

\bibitem{FT} L.Faddeev, L.Takhtajian, Russian Math. Surveys {\bf 34:5}
  (1979) 11\\
  V.Korepin, N.Bogoliubov, A.Izergin, \textsl{Quantum Inverse
  Scattering Method
  and Correlation Functions,}  Cambridge Univ.Press (1993).

\bibitem{W} Y.Umeno, M.Shiroishi and M.Wadati, J.Phys.Soc.Jpn.
{\bf 67} (1998) 2242.

\bibitem{AKMS} J.Ambjorn,D.Karakhanyan,M.Mirumyan,A.Sedrakyan,
    Nucl.Phys.{\bf 599 B}(2001)547.

\bibitem{F} F.~Berezin,  The Method of Second Quantization(Nauka,
  Moscow 1965), \\
  L.~Faddeev, Introduction to Functional Methods, in Les Houches(1975)\\
  Session 28, ed. R.~Balian, J.~Zinn-Justin.

\end{thebibliography}
\end{document}